\documentclass[11pt,draftcls,onecolumn]{IEEEtran}
\pdfoutput=1
\usepackage{graphicx}
\usepackage{color}
\usepackage{amsmath}
\usepackage{amsfonts}
\usepackage{amssymb}
\usepackage{mathtools}
\usepackage[english]{babel}
\usepackage[latin1]{inputenc}
\usepackage{cite}
\usepackage{algorithm}
\usepackage{enumitem}

\usepackage{stfloats} 
\usepackage{multirow}
\usepackage{caption}
\usepackage{subcaption}
\usepackage{wrapfig}
\usepackage{lscape}
\usepackage{rotating}

\newcommand{\calN}{{\cal N}}

\newcommand{\iid}{\stackrel{iid}{\sim}}

\graphicspath{{./}{./figures/}}

\newtheorem{rem}{Remark}
\begin{document}

 \title{Noise Robust Online Inference for Linear Dynamic Systems}
\author{Saikat~Saha 
    \thanks{ S. Saha is with the
Division of Automatic Control, Department of Electrical Engineering, Link\"{o}ping University, 
Sweden, \texttt{saha@isy.liu.se}}
}

\date{}

\maketitle

\begin{abstract}
We revisit the Bayesian online inference problems for the linear dynamic systems (LDS) under non-Gaussian environment. The noises can naturally be non-Gaussian (skewed and/or heavy tailed) or to accommodate spurious observations, noises can be modeled as heavy tailed. However, at the cost of such noise robustness, the performance may degrade when such spurious observations are absent. Therefore, any inference engine should  not only be robust to noise outlier, but also be adaptive to potentially unknown and time varying noise parameters; yet it should be scalable and easy to implement. 

To address them,  we envisage here a new noise adaptive Rao-Blackwellized particle filter (RBPF), by leveraging a hierarchically Gaussian model as a proxy for any non-Gaussian (process or measurement) noise density. This leads to a  conditionally linear Gaussian model (CLGM), that is tractable. However, this framework requires a valid transition kernel for the intractable state, targeted by the particle filter (PF). This is typically unknown. We outline how such kernel can be constructed provably, at least for certain classes encompassing many commonly occurring non-Gaussian noises, using auxiliary latent variable approach. The efficacy of this RBPF algorithm is demonstrated through numerical studies.

\end{abstract}

%
%
%
%
%
%
%
\section{Introduction}
\noindent
Many applications of interest to the signal processing community (e.g., tracking, autonomous navigation and surveillance) require that the inference must be performed on near real-time (online) using the streaming sensor data. 
The stability and the performance of  the underlying inference engines however depend crucially on the sensor data  quality. Usually, any spurious sensor data needs to be detected and discarded before passing to the inference engine, so that the latter is not susceptible to permanent failure. In recent times, the sensors are becoming cheaper, however at the expense of their performance reliability. The proliferation of such inexpensive sensors has opened up the possibility to explore many complex and high dimensional problems (e.g., motion tracking, road traffic monitoring), which are hitherto impossible with a limited number of costly sensors. This trend for using inexpensive sensors has in turn, laid greater emphasis on the processing algorithms to the extent that any inference algorithm (presumably with more computing prowess) is required to be robust and stable against such spurious sensor data; yet simple to implement and vastly scalable to the high dimensional problems. Against this backdrop, we consider the online inference problems for the discrete time LDS. 

 

\subsection{Problem Background}
Consider the following discrete time LDS relating the latent state $x_{k}\in \mathbb{R}^{n_x}$ at time step $k$ to the observation $y_{k}\in \mathbb{R}^{n_y}$ as 
 \begin{subequations}
  \label{eq:LDS}
  \begin{align}
    x_{k} &=  A_{k}\ x_{k-1}+ B_{k}\ w_k,\\ %
    y_{k} &=  C_{k}\ x_{k} + e_{k},
  \end{align}
\end{subequations}
 where $w_k$ and $e_k$ are the process and measurement noise respectively. The noises are assumed to be independent and also independent of each other. The model parameters $\{A_{k},B_{k},C_{k}\}$ are considered to be known here. Given this model, an initial state prior (i.e., $p(x_0)$) and a stream of observations up to time $k$, $y_{0:k}\triangleq \{y_0,y_1,\dots,y_k\}$, one typical inference task is to optimally estimate the sequence of (posterior) densities $p(x_{k}|y_{1:k})$, in an online fashion over time. This is known as the online Bayesian filtering problem for the LDS, and the density $p(x_{k}|y_{1:k})$ is called as the filtering density. When the noises above are Gaussian, the filter density can be obtained recursively in closed form using the celebrated Kalman filter (KF)\cite{West_h97,Bagchi,Anderson_79}. However, this analytical tractability is lost if any noise deviates from such nominal Gaussian assumptions. 

In reality, many noise sources naturally appear to be non-Gaussian (characterized by their heavy tails and/or skewness). For example, noise with an impulsive nature (sharp spikes and/or occasional bursts) appears in many applications such as speech and audio signal processing, astrophysical imaging, underwater navigations, multi-user radar communications, kick detection in oil drilling, finance and insurance among others \cite{Chenssp14,Jordi14,ZoubirSPM,Mila_stable,Hawary95,Godsill96,Fruhwirth07,Adler98}. This impulsive nature can  be modeled e.g., by a noise distribution that has heavier tails than the Gaussian distribution. Heavy tailed distributions are also used to model the presence of so called \textit{outliers}, which are data points, that do not appear to follow the pattern of the other data points \cite{Ross}. Data from  the visual sensors, GPS devices, sonar, and radar are often contaminated by such outliers. The root causes of such outlying observations are often unknown or are excluded deliberately from the model due to the complexity and the computational issues. So under such simplified modeling of complex real world processes, these unusual observations can be taken care by noise outliers. This in turn, requires heavy tailed distributions for the noises\cite{Pearson02}.

In the filtering context,  when a nominal model with specified Gaussian noises cannot account for the outliers or sudden change in unknown input signals, the filter becomes unstable. Since for the real world applications, often we have  weak knowledge on the systems and outliers are frequent,  naturally, the filtering problems under heavy tailed noises have attracted considerable research attentions \cite{Agamennoni11, Piche_SS_JH12, Roth_Lic, Ting_ecml07}. We note that in the context of filtering, the process and/or the observation noise can have heavy tails. Although heavy tailed observation noise have received  much attention, in applications like maneuvering target tracking, modeling occasional pilot induced changes requires a heavy tailed process noise \cite{Sinha_TAES}. 
In contrast, although skewness appears in many applications \cite{Kok_Lic}, the associated online inference problem has not been well explored \footnote{One notable exception is the recent article \cite{Nurminenetal:15}, that has been brought to our notice.}.

Non-Gaussian noise can also appear due to the modeling artifact. We illustrate the last point through the following example: 
\subsubsection{Stochastic volatility model}\label{mot:ex1}
We consider the following discrete time model \cite {Shephard96}
\begin{subequations}\label{SVmodel}
\begin{align}
h_{k+1}&= \gamma_0 + \gamma_1 h_k + \eta_k,\\
y_k &=  \epsilon_k  exp(h_{k}/2),
\end{align}
\end{subequations}
where $h_k$ and $y_k$ are the latent log-volatility and observed asset return at time step $k$ with $|\gamma_1|<1$, $h_0 \sim \mathcal N(0,\frac{\sigma_n^2}{1-\gamma_1^2})$, $\eta_k \iid \mathcal N(0,\sigma_n^2)$ and $\epsilon_k\iid \mathcal N(0,1)$, where \textit{iid} means independent and identically distributed. $\gamma_0, \gamma_1$ and $\sigma_n$ are assumed to be known.
Here  $\eta_k$ is uncorrelated to $\epsilon_k$, i.e. no leverage effect is considered. Note that the measurement noise is multiplicative. The above dynamic model can equivalently be cast as a linear state space model:
\begin{subequations}
\begin{align}
h_{k+1}&=  \gamma_0 +\gamma_1 h_k + \eta_k, \label{LSV:dyn}\\
log(y_k^2)&=  h_k + log(\epsilon_k^2)\label{LSV:obs}.
\end{align}
\end{subequations}
However the observation noise $e_k \equiv log(\epsilon_k^2)$, being log of a $\chi^2$ random variable, is no more Gaussian. In fact, the density of this noise is available analytically, given by (see e.g.,\cite{Douc14})
\begin{equation}\label{vol_skew}
f(e_k) = \frac{1}{\sqrt{2\pi}}{exp\big\{-0.5\big(exp(e_k)-e_k\big)\big\}},\ -\infty<e_k< \infty.
\end{equation}
\begin{figure}[ptbh]
  \centering
	\includegraphics[width=2.7in, height= 1.4in]{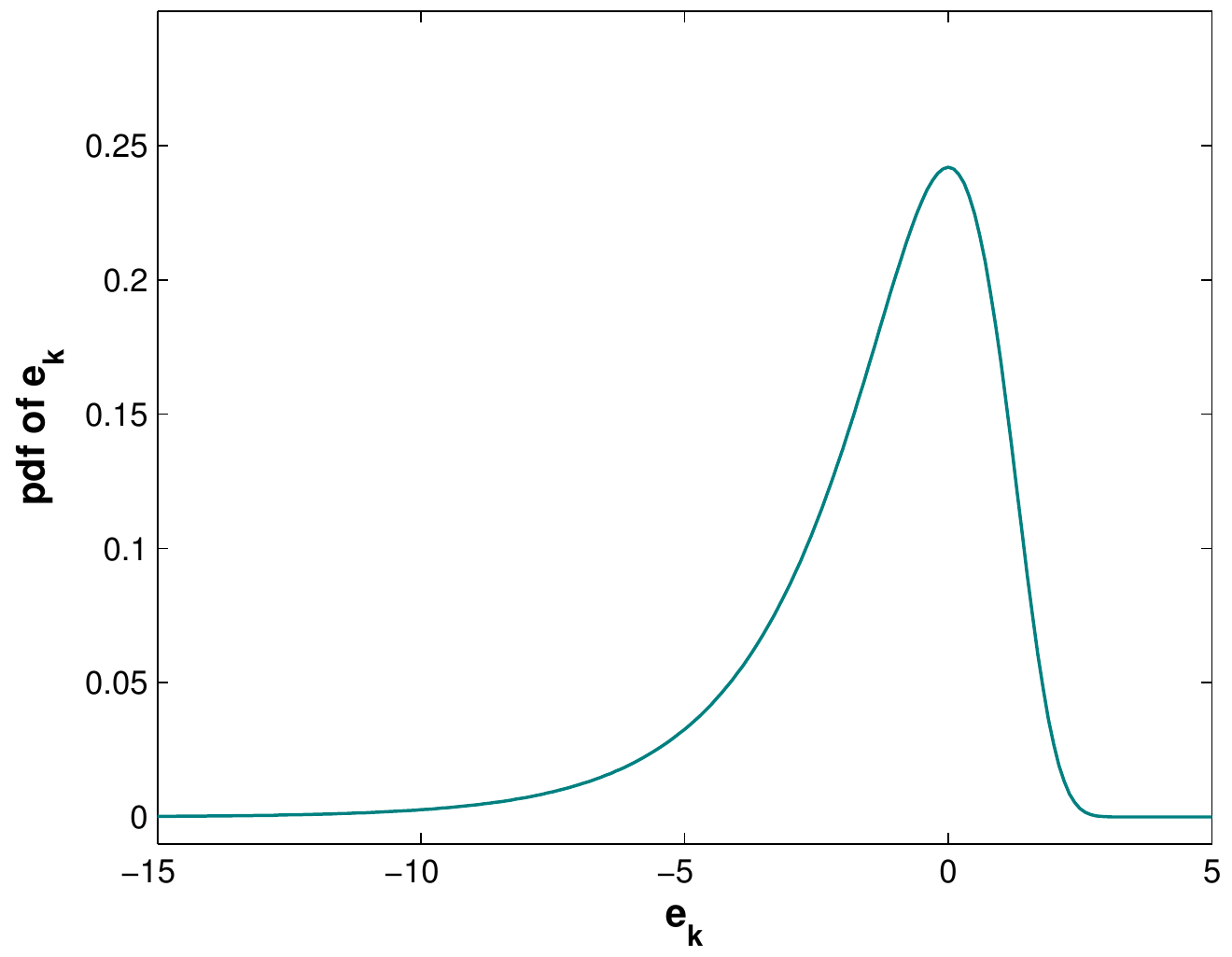}
  \caption{Density plot for the observation noise in \eqref{vol_skew}.}
  \label{fig:Skewd_noise}
\end{figure}
From Figure (\ref{fig:Skewd_noise}), it is evident that the density is highly skewed. Thus, we have a LDS driven by a non-Gaussian (measurement) noise.\\ 

While the noise distribution is certainly a key factor for the inference task, the other equally important aspect for practical considerations is the knowledge on the noise statistics. In many practical contexts, although prior knowledge on the noise distributions is fairly available, the corresponding noise parameters are often unknown. Thus additionally, the noise parameters need to be estimated on the fly as well. This is known as online noise adaptive filtering. When the noise parameters are stationary, the online Bayesian estimation of the parameters is known to be difficult and one practical solution is to assume the noise parameters to be slowly varying in time\cite{OzkanSSLG13}. However, this assumption easily breaks down in the presence of any potential outlier. To get rid of the outliers, some mechanisms are usually placed to detect and immediately discard them.
However such outlying observations may carry important information about e.g., any unmodeled system characteristics and model degradation   
and as such, as pointed out in \cite{Cappe05}, "\texttt{Routinely ignoring unusual observations is neither wise nor statistically sound}". On the other hand, eventhough such outliers can be accommodated by using e.g., a properly specified heavy tailed noise distribution from a known parametric family, the performance of the filter can be degraded severely  when the outliers are absent. This happens due to the use of fixed form of distribution.  To alleviate this problem in noise adaptive filtering, essentially what we need is to use a heavy tailed noise, whose parameters are time varying. However, coping with  
nonstationary noise parameters in online setting is very challenging as the corresponding dynamics for the parameters cannot be easily specified \textit{a priori} in practice.

\subsection{Prior works}
There is a long history on the efforts of improving the KF algorithm under the non-Gaussian environments. The earlier attempts were mainly based on the  robustification arguments in the presence of outliers.  These were primarily addressed either by analytical approximations using elliptical noise distributions or by heuristic cost functions in the update step of KF. The elliptical noise based approaches \cite{Meinhold89,Giron94} were not robust, as the posterior mean is unbounded as a function of the residuals, whereas  the approaches based on the ad-hoc cost functions (e.g., \cite{Masreliez75,Durovic}) requires tuning of parameters and their implementations were very involved as compared to KF. Later, the simulation based approaches (e.g., PF)\cite{DelMoral04,Cappe05} became popular due to their ease of implementations. However, the major disadvantages were their non-scalability together with computational costs\cite{Snyder08}. Several recent articles have addressed the filtering problem using variational Bayes (VB) \cite{Ting_ecml07,Agamennoni11,Piche_SS_JH12} and convex optimization based methods \cite{Boyd}.
Although VB method is  quite scalable,  in general, it requires fairly involved mathematical derivations and  is known to consistently underestimate the posterior covariance.
 Common to all these recent studies is the assumption of outliers in the observation noise only. The methods by \cite{Ting_ecml07,Piche_SS_JH12, Boyd} do not accommodate any persistence (time correlated) noise outliers.
Moreover, \cite{Piche_SS_JH12} considers a heuristic transition model for the noise parameter (as observed by \cite{Agamennoni11}, the filter is not stable under abrupt change in noise) whereas \cite{Agamennoni11,Boyd} require additional input parameter from the user.
More recently, observing that the VB based algorithms are very sensitive to process noise outlier,  \cite{Roth_Lic} proposed  an analytical approximation based \textit{t} filter, where both the process and observation noises are Students' \textit{t}. The approximations here require that the estimated state and process/observation noise is jointly \textit{t} with a common degree of freedom (dof) parameter. To maintain these hypotheses and to prevent
the growth of dof parameters in filter recursion, again careful tuning is required at each step. For all the cases, the noise class is implicitly assumed to be symmetric and Students' \textit{t}.  

From the literature above, we see that many of the existing approaches require manual tuning of the parameters and also their implementations are substantially involved as compared to well understood KF. Thus any future inference framework should ideally be  free of such heuristic tuning and the implementation should be simple to understand. Also, any future framework should be able to handle-(a) asymmetric noise and (b) symmetric heavy tailed noise other than Students'  \textit{t}.  

\subsection{Sketch of our contributions}
Keeping in view the points above, we consider the online inference for the LDS under a fairly general and realistic scenario. This is addressed here in two stages by successively dropping the following common assumptions: (a) noises are Gaussian and (b) noise parameters are known. 

In the first stage, we consider the LDS under (known) stationary non-Gaussian environment. The corresponding inference task is analytically intractable. In our approach, we envisage a hierarchically Gaussian model (HGM) as a proxy for any non-Gaussian noise. This model can represent the skewness and/or heavy tail that are typical characteristics of the non-Gaussian noise. Such  representation allows us to exploit a CLGM, which is analytically tractable using a KF. This in turn, leads to a  RBPF framework for the online inference task. Since within the RBPF framework, PF is  confined to a space of lower dimension, it acts as an enabler for scaling to high dimensional problems. The proposed framework here uses a bank of KFs, which is simple to understand; the  sophistication comes in the way how we mix and propagate the output of different KFs to get the target filter density. Although this RBPF framework is not entirely new \cite{chen:liu:00:mixture}, a proper specification of the transition kernel for the state, targeted by PF is often difficult and this technical aspect has received little attention. We elaborate this issue in details and address the proper transition kernel for certain class of noises (where widely used Students' \textit{t} is a special case). 

For the second stage, we point how this RBPF framework is already doing a noise adaptive filtering. However to make the filter robust against any large noise deviation, we need prior knowledge about such deviation. Often this is reasonably well known for many practical applications. For such cases, we outline how this information can be encoded in the noise prior. This completes our noise robust online inference framework.
%
%
%
%
\subsection{Organization of the article}
\noindent
The rest of the article is organized as follows. We start with brief descriptions on HGM in section \ref{Hier_gauss} and  simulation based online filtering in section \ref{sim_filt}. This is then followed by the development of the proposed inference framework for LDS under stationary non-Gaussian environment in section \ref{LDS_inf}. We then describe how the above framework can be used for robust noise adaptive filtering in section \ref{rob_noise_adap}. Subsequently, the numerical experiments are shown in \ref{num_experiment}, which is followed by concluding remarks in section \ref{conclusion}.
%
%
%
%
%
%
%
%
%
%
%
%
%

%
%
%
%
%
%
\section{Hierarchical Gaussian model (HGM) for non-Gaussian density}\label{Hier_gauss}
\noindent
Non-Gaussian densities are often characterized by the presence of heavy tail and/or skewness. Many such densities can often be modeled as hierarchically Gaussian. That is, the density can be represented as Gaussian, conditionally on an auxiliary random variable, known as the 'mixing parameter'. We outline below a brief description for such hierarchical Gaussian representation. For the clarity of the presentation, we consider two separate classes based on the symmetry of underlying density.
When the density is symmetric, HGM is represented as scale mixture of normal (SMN), also known as normal variance mixture (NVM). For the non symmetric density, it is represented as normal variance-mean mixture (NVMM) model. 
%
%
%
\subsection{Scale mixtures of  normal}\label{NVM}
A random vector $X$ has a scale mixtures of  normal density, if it can be expressed as follows
\begin{align}
	X = \mu + \kappa^{\frac{1}{2}}(\Lambda)Z,
\end{align}
where  $\mu$ is a location vector, $Z$ is distributed according to a zero mean multivariate normal with covariance matrix $\Sigma$ and $\kappa$ is a positive weight function. $\Lambda$ is a random variable, known as mixing parameter and is distributed on the positive real line, independent of $Z$. Note here that conditioned on $\Lambda = \lambda$,  $X$ follows a multivariate normal distribution with mean vector $\mu$ and covariance matrix $\kappa(\lambda)\Sigma$, i.e., $(X|\Lambda = \lambda)\sim \calN(x|\mu, \kappa(\lambda)\Sigma)$. Then, the probability density function (pdf) of $X$ can be expressed as
\begin{equation}\label{SMN_int_rep}
	p(x) = \int_{0}^{\infty} \calN (x|\mu, \kappa(\lambda)\Sigma)p(\lambda)d\lambda,
\end{equation}
where $p(\lambda)$ is the density function of $\Lambda$. $p(\lambda)$ is referred to as the mixing density of SMN representation.

The symmetrical class of SMN includes among others, the popular Students' \textit{t}, the Pearson type VII family, the Slash and the variance gamma distributions\footnote{The Gaussian distribution in this context, can be thought as a degenerate mixture \cite{West87}}. All these distributions are characterized by their heavy tails as compared to the normal distribution. Although there exists many other important SMN distributions (e.g., exponential power, symmetric $\alpha$ -stable, logistic, horse shoe, symmetric generalized hyperbolic distribution)\cite{ChoyChan08,carvalho2010}, in the sequel, we present only the above mentioned special cases, as the associated mixing densities have computationally attractive form.

\subsubsection{Multivariate Students' \textit{t} distribution}
When $X$ follows a multivariate Students' \textit{t} distribution with location $\mu$, scale $\Sigma$ and degrees of freedom $\nu$ (i.e., $X \sim \mathcal{T}(\mu,\Sigma;\nu)$), the pdf of $X$ can be expressed in the following SMN form
\begin{equation}
p(x) = \int_{0}^{\infty} \calN (x|\mu, \frac{\Sigma}{\lambda}) Ga(\lambda|\frac{\nu}{2}, \frac{\nu}{2})d\lambda,
\end{equation}
where $Ga(\cdot|a,b)$ is the gamma density function of the form
\begin{equation}
\label{Gammadensity}
Ga(\lambda|a,b)= \frac{b^{a}}{\Gamma(a)}\lambda^{a-1}e^{-b\lambda}, \ \ \lambda,a,b > 0,
\end{equation}
 and $\Gamma(a)$ is the gamma function with argument $a>0$. Consequently, $X$ can be presented in the following hierarchical form
\begin{equation}
X|(\mu,\Sigma,\nu, \lambda) \sim \calN(\mu,\lambda^{-1}\Sigma), \ \ \lambda|\nu \sim Ga(\frac{\nu}{2}, \frac{\nu}{2}). 
\end{equation}

\subsubsection{Pearson type VII distribution}
If $X$ belongs to the Pearson type VII family,  the associated density is given by
\begin{equation}
p(x|\mu,\Sigma,\nu,\delta)= \frac{1}{B(\frac{\delta}{2}, \frac{1}{2})\sqrt{\nu\Sigma}}\big(1+\frac{(x-\mu)^2}{\nu\Sigma} \big)^{-(\delta+1)/2},
\end{equation}
where $\mu$ and $\Sigma$ are the location and scale parameters, $\nu>0$ and $\delta>0$ are the shape parameters and $B(a,b)$ is the beta function with arguments $a>0$ and $b>0$. The Pearson type VII density can be expressed hierarchically as
\begin{equation}
X|(\mu,\Sigma,\nu,\delta,\lambda) \sim \calN(\mu,\lambda^{-1}\Sigma),\ \ {\lambda}|(\nu,\delta) \sim Ga(\frac{\nu}{2}, \frac{\delta}{2}).
\end{equation}
We get back to Students't distribution when $\nu=\delta$ and Cauchy if $\nu=\delta=1$.

\subsubsection{Multivariate slash distribution}
The Slash distribution can be hierarchically represented as
\begin{equation}
X|(\mu,\Sigma,\nu) \sim \calN(\mu,\lambda^{-1}\Sigma), \ \ \lambda|\nu \sim Be(\nu,1),
\end{equation}
with $0< \lambda<1$ and $\nu>0$, where $Be(\cdot)$ denotes the Beta distribution.
\subsubsection{Multivariate variance gamma distribution}\label{sec:vg}
When $X$ follows a multivariate variance gamma (VG) distribution with location $\mu$, scale $\Sigma$ and degrees of freedom $\nu>0$ (i.e., $X \sim \mathcal{VG}(\mu,\Sigma;\nu)$), the density can be represented in the following hierarchical form
\begin{equation}\label{VG-hier}
X|(\mu,\Sigma,\nu, \lambda) \sim \calN(\mu,\lambda^{-1}\Sigma), \ \ \lambda|\nu \sim IG(\frac{\nu}{2}, \frac{\nu}{2}), 
\end{equation}
where $IG(a,b)$ is the inverse gamma density given by
\begin{equation}
\label{IGdensity}
IG(\lambda|a,b)= \frac{b^{a}}{\Gamma(a)}\lambda^{-(a+1)}e^{-b/\lambda}. 
\end{equation}
When $\nu=2$, VG  turns into a Laplace distribution.

\subsection{Normal variance-mean mixture}\label{NVMM}
A random vector $X$ is following a normal variance-mean mixture distribution, if it can be expressed as follows \cite{Nielsen82}
\begin{align}
\label{GHdist}
	X = \mu + \beta \Lambda + \sqrt{\Lambda}Z,
\end{align}
where $\Lambda$ is independent of $Z$ and $Z$ is distributed according to a zero mean multivariate normal with covariance matrix $\Sigma$. The random variable $\Lambda$ is the mixing parameter and is distributed on the positive real line and $\mu, \beta \in \mathbb{R}$. The distribution of  $X$ conditioned on $\Lambda = \lambda$ is multivariate normal given as
$(X|\Lambda = \lambda)\sim \calN(x|\mu + \beta \lambda, \Sigma \lambda)$. Note that when $\beta=0$, the NVMM turns into a SMN with $\kappa(\lambda)=\lambda$. We describe below some special cases of this class of distributions:

\subsubsection{Generalized hyperbolic distribution}
$X$ in \eqref{GHdist} follows a generalized hyperbolic (GH) distribution, if $\Lambda$ is distributed according to a generalized inverse Gaussian (GIG) distribution as
\begin{equation}
p(\lambda) = \frac{(a/b)^{p/2}}{2K_{p}(\sqrt{ab})}\lambda^{p-1}exp\{-\frac{a\lambda+b/\lambda}{2}\}, \ \lambda > 0, 
\end{equation}
where $K_{p}$ is  the modified Bessel function of the second kind, $a,b \in \mathbb{R}^{+}$,  and $p \in \mathbb{R}$.

\subsubsection{GH skew Students' \textit{t} distribution}
The hierarchical structure of GH skew Students' \textit{t} distribution is given as \cite{Nakajima:csda12}
 \begin{equation}
(X|\Lambda = \lambda)\sim \calN(x|\mu + \beta \lambda, \Sigma \lambda), \
\lambda|\nu  \sim IG(\nu/2,\nu/2),
 \end{equation}
where inverse-gamma density is given by \eqref{IGdensity}. Note that the GH skew Students' \textit{t} is a special case of the GH distribution, where   the parameters for the GIG are selected as $a=-\nu/2 \ (\nu>0)$, $b=\sqrt{\nu}$ and $p = 0$. Moreover, when $\beta = 0$, this turns to a symmetric Students' \textit{t} distribution and if $\nu \rightarrow \infty$, this becomes a skew normal distribution.

\subsubsection{GH variance gamma distribution}
The hierarchical structure of GH variance gamma distribution is  given as
 \begin{equation}
(X|\Lambda = \lambda)\sim \calN(x|\mu + \beta \lambda, \Sigma \lambda), \
\lambda|\nu  \sim Ga(\nu/2,\nu/2),
 \end{equation}
where the gamma density is given by \eqref{Gammadensity}.

\rem\label{remark2} note that the mixing parameter $\Lambda$ in \ref{NVM}  and \ref{NVMM} being a random variable, dimension of $\Lambda$ is always less than or equal to that of random vector $X$. This observation is important, as we will see later that the PF can be used to target the low dimensional $\Lambda$ in a RBPF framework.

\section{Simulation based online filtering}\label{sim_filt}
\noindent
When a dynamic system (state space model) is nonlinear and/or driven by non-Gaussian noises, the filter density $p(x_{k}|y_{1:k})$ is in general, analytically intractable. To deal with, many approximated methods have been proposed over time \cite{Murphy_book}. Particle filtering (PF) is one such method, which uses Monte Carlo simulations to address the filtering problem. In this section, we give a very brief overview of PF and RBPF.

\subsection{Particle filtering (PF)}
In PF, the posterior distribution associated with the density $p(x_{0:k}
|y_{1:k})$ is approximated by an empirical distribution induced by a
set of $N (\gg 1)$ weighted particles (samples) as
\begin{equation}
  \widehat{P}_{N}(dx_{0:k}|y_{1:k})
  = \sum_{i=1}^{N}\widetilde{w}_{k}^{(i)}\delta_{x_{0:k}^{(i)}}(dx_{0:k}); \ \ \widetilde{w}_{k}^{(i)} \ge 0, \label{PF_cloud}
\end{equation}
where $\delta_{x_{0:k}^{(i)}}(A)$ is a Dirac measure for a given
$x_{0:k}^{(i)}$ and a measurable set~$A$, and
$\widetilde{w}_{k}^{(i)}$ is the associated weight attached to each
particle $x_{0:k}^{(i)}$, such that
\mbox{$\sum_{i=1}^{N}\widetilde{w}_{k}^{(i)}=1$}. 
Even though the distribution $\widehat{P}_{N}(dx_{0:k}|y_{1:k})$ does
not admit a well defined density with respect to the Lebesgue measure, we use notational abuse to represent the associated empirical density as
\begin{equation}
  \label{PF:density}
  \widehat{p}_{N}(x_{0:k}|y_{1:k})
  = \sum_{i=1}^{N}\widetilde{w}_{k}^{(i)}\delta(x_{0:k}-{x_{0:k}^{(i)}}).
\end{equation}
Although \eqref{PF:density} is not mathematically rigorous, it is intuitively easier to follow than the stringent measure theoretic notations, especially if we are not concerned with the theoretical convergence studies.

Note that the posterior $p(x_{0:k}|y_{1:k})$, which we target using a PF, is unknown. The empirical distribution  $\widehat{P}_{N}(dx_{0:k}|y_{1:k})$ in \eqref{PF_cloud} is obtained by first generating samples $x_{0:k}^{(i)}$ from a proposal distribution $\pi(x_{0:k}|y_{1:k})$ and then the corresponding weights are obtained using the idea of importance sampling as
\begin{equation}
{w}_{k}^{(i)} = \frac{p(x_{0:k}^{(i)}|y_{1:k})}{\pi(x_{0:k}^{(i)}|y_{1:k})}; \ \widetilde{w}_{k}^{(i)}
  = \frac{{w}_{k}^{(i)}}{\sum_{j=1}^{N}{w}_{k}^{(j)}}.
\end{equation}
Given this PF
output, one can now approximate the marginal distribution $p(x_{k}|y_{1:k})$ as
  $\widehat{P}_{N}(dx_{k}|y_{1:k})
  = \sum_{i=1}^{N}\widetilde{w}_{k}^{(i)}\delta_{x_{k}^{(i)}}(dx_{k})$.
Suppose at time $(k-1)$, we have a weighted particle approximation
of the posterior $p(x_{0:k-1}|y_{1:k-1})$ as
$\widehat{P}_{N}(dx_{0:k-1}|y_{1:k-1})
=\sum_{i=1}^{N}\widetilde{w}_{k-1}^{(i)}\delta_{x_{0:k-1}^{(i)}}(dx_{0:k-1})$. Now with a new measurement $y_{k}$, we wish to approximate
$p(x_{0:k}|y_{1:k})$ with a new set of particles (samples). 
A standard PF uses the following posterior path-space recursion
\begin{equation}\label{PF_gen_recursion}
p(x_{0:k}|y_{1:k}) \propto  p(y_{k}|x_{0:k},y_{1:k-1}) 
  p(x_{k}|x_{0:k-1},y_{1:k-1})\,p(x_{0:k-1}|y_{1:k-1}). 
\end{equation}
Now for the Markovian state space model, this becomes
\begin{equation}\label{PF_markov_rec}
 p(x_{0:k}|y_{1:k}) \propto  p(y_{k}|x_{k})\,
  p(x_{k}|x_{k-1})\,p(x_{0:k-1}|y_{1:k-1}).		
\end{equation}
Assuming that the proposal distribution can be decomposed as 
$\pi(x_{0:k}|y_{1:k}) = \pi(x_{k}|x_{0:k-1},y_{1:k})\pi(x_{0:k-1}|y_{1:k-1})$,
we can now implement a sequential importance sampling,  where the particles are
propagated to time $k$ by sampling a new state $x_{k}^{(i)}$ from the marginal
proposal kernel $\pi(x_{k}|x_{0:k-1}^{(i)},y_{1:k})$ and setting
$x_{0:k}^{(i)} \triangleq \left(x_{0:k-1}^{(i)}, x_k^{(i)}\right)$.
Subsequently using \eqref{PF_markov_rec}, the corresponding weights of the particles can be given by
\begin{align}
  \label{WT_rec}
  w_{k}^{(i)}
  &\propto 
  \frac{p(y_{k}|x_{k}^{(i)})p(x_{k}^{(i)}|x_{k-1}^{(i)})}
  {\pi(x_{k}^{(i)}|x_{0:k-1}^{(i)},y_{1:k})}\widetilde{w}_{k-1}^{(i)}; \ \
  \widetilde{w}_{k}^{(i)}
 = \frac{{w}_{k}^{(i)}}{\sum_{j=1}^{N}{w}_{k}^{(j)}}.
\end{align}
 
To avoid carrying trajectories with small weights and to concentrate
upon the ones with large weights, the particles need to be resampled
regularly. The effective sample size $N_{eff}$, a measure of how many particles that actually contributes to the approximation of the distribution, is often used to decide when to resample. When $N_{eff}$ drops below a specified
threshold, resampling is performed. Many efficient resampling schemes have been proposed in the literature. Instead of going into the details, we refer the interested readers to \cite{DelMoral04,Doucet:Johansen11,Cappe05,Gustafsson:10,Djuric_SPM_PF} for a more general introduction to PF. 
\rem If one uses the state transition density as proposal with resampling at each step, the corresponding PF is known as  \textit{bootstrap particle filter}. This is easy to implement, so very popular in practice.

\subsection{Rao-Blackwellized particle filtering (RBPF)}
Although PF is very popular and has been around for a while, it is computationally demanding and notably, it has severe limitations when scaling to
higher dimensions \cite{Snyder08}. For certain models, when part of the state space is (conditionally) tractable, it is then sufficient to employ a PF for the remaining intractable part of the state space. If it is possible to exploit such analytical substructure, the Monte Carlo based estimation is then confined to a space of lower dimension. As a consequence, the estimate obtained is often better (in terms of asymptotic variance) and never worse than the estimate provided by the PF targeting the full state space. The resulting method is popularly known as \emph{Rao-Blackwellized particle filtering}\cite{chen:liu:00:mixture, Chopin04, Schon05, Saha_RBPF_rev14,SahaOGS09}. 
Note that solving part of the state vector analytically (or using analytical approximations) leaves the remaining part to be targeted only by PF. Thus RBPF acts as a practical enabler for scaling to high dimensional problems.


%
%
%
%
%
%
%
\section{Inference for LDS under stationary non-Gaussian environment}\label{LDS_inf}
\noindent
Consider the LDS in \eqref{eq:LDS}, driven by  known stationary non-Gaussian noises,  assumed to be hierarchically Gaussian as in Section \ref{Hier_gauss}. Given the model and assuming that the initial prior, $p(x_{0})$ is a known Gaussian density\footnote{More generally, $p(x_{0})$ can be taken to be a hierarchically Gaussian density.}, our objective is to recursively target the intractable filtering density $p(x_{k}|y_{1:k})$. 

Let at time step $k$, $\lambda_{k}^{w}$ and $\lambda_{k}^{e}$ be the corresponding mixing parameters  for $w_k$ and $e_k$. We define an auxiliary vector $\lambda_{k} \equiv (\lambda_{k}^{w},\ \lambda_{k}^{e})^T$. Then the noises $w_k$ and $e_k$ are hierarchically Gaussian as  
\begin{subequations}\label{cg_noises}
\begin{equation}
	w_k|\lambda_{k} \sim  \mathcal N\big(\mu^{w}_{k}(\lambda_{k}), Q_{k}(\lambda_{k})\big),
	\end{equation}
	\begin{equation}
	e_{k}|\lambda_{k} \sim \mathcal N\big(\mu^{e}_{k}(\lambda_{k}), R_{k}(\lambda_{k})\big),
\end{equation}
\end{subequations}
where $\mu^{w}_{k}$, $\mu^{e}_{k}$ are the corresponding mean and $Q_{k}$ and $R_{k}$ are the corresponding covariance of the hierarchical Gaussian noises\footnote{Instead of both the noises being non-Gaussian, if one of them is Gaussian, the inference framework, described subsequently, is still valid. In that case $\lambda_k$ becomes an auxiliary random variable, representing the mixing parameter of the non-Gaussian noise.
}. These parameters  
can possibly depend upon $\lambda_{k}$. Such a noise representation admits a CLGM, which is analytically tractable using the KF. This opens up a RBPF implementation for the online inference problem. In the sequel, we describe this RBPF framework. This section is organized as follows.
We first outline one complete iteration of the proposed RBPF framework, which is followed by the specification of the dynamics for the state (i.e., mixing parameters), targeted by associated PF. Next, an algorithmic description of the approach is provided. This is then followed by the descriptions on the likelihood function estimation and  p-step ahead prediction. 

\subsection{Posterior propagation cycle of RBPF}
At time zero,  $p(x_0)$ is known. 
Suppose at time step $(k-1)$, we have the joint target distribution $p(x_{k-1},\lambda_{1:k-1}|y_{1:k-1})$. This can be decomposed as 
\begin{equation}\label{factor_jt_pos}
p\biggl({x_{k-1}, \atop \lambda_{1:k-1}} \biggm|{y_{1:k-1}}\biggr)
= p\biggl(x_{k-1}\biggm|{\lambda_{1:k-1}, \atop y_{1:k-1}}\biggr)\ p\biggl(\lambda_{1:k-1}|y_{1:k-1}\biggr),	
\end{equation}
where $p(\lambda_{1:k-1}|y_{1:k-1})$ is targeted by a PF and is empirically given by a set of $N(>>1)$ weighted random particles as
\begin{align}\label{PF_output}
p(\lambda_{1:k-1}|y_{1:k-1})
  = \sum_{i=1}^{N}\widetilde{w}_{k-1}^{(i)}\delta(\lambda_{1:k-1}-\lambda_{1:k-1}^{(i)}),
\end{align}
with $\widetilde{w}_{k-1}^{(i)} \ge 0$ and $\sum_{i=1}^{N}\widetilde{w}_{k-1}^{(i)}=1$. Using \eqref{PF_output}, we have
\begin{align}\label{PF_marginal}
p(\lambda_{k-1}|y_{1:k-1})
  = \sum_{i=1}^{N}\widetilde{w}_{k-1}^{(i)}\delta(\lambda_{k-1}-\lambda_{k-1}^{(i)}).
\end{align}
Noting that, given a sequence $\lambda_{1:k-1}^{(i)}$, the dynamic system now becomes a CLGM. So $p(x_{k-1}|\lambda_{1:k-1}^{(i)},y_{1:k-1})$ can be obtained by a KF,
 given by
\begin{align}\label{KF_output}
	p(x_{k-1}|\lambda_{1:k-1}^{(i)},y_{1:k-1}) = \mathcal N \big(\widehat{x}_{k-1}(i), P_{k-1}(i)\big),
\end{align}
where for notational clarity, we suppress the dependency of the parameters on $\lambda_{1:k-1}^{(i)}$. Since a KF is running for each sequence (henceforth denoted by an index $i$), we have total $N$ number of KFs running in parallel at any given time.
Now using \eqref{PF_output} and \eqref{KF_output} together, 
the filter distribution $p(x_{k-1}|y_{1:k-1})$ can be obtained as
\begin{eqnarray}\label{filter_target}
p(x_{k-1}|y_{1:k-1})&=& \int p\biggl(x_{k-1} \biggm|{\lambda_{1:k-1}, \atop y_{1:k-1}}\biggr) \nonumber \\
& & \times \ p\biggl(\lambda_{1:k-1}\biggm|y_{1:k-1}\biggr) d\lambda_{1:k-1}, \nonumber\\
	&\approx &\sum_{i=1}^{N}\widetilde{w}_{k-1}^{(i)}\ \mathcal N\biggl(\widehat{x}_{k-1}(i), P_{k-1}(i)\biggr),
\end{eqnarray}
which is a weighted (finite) mixture of $N$  Gaussian distributions. The mean and covariance of this filter distribution (assuming them to be finite) can be given as
\begin{subequations}	
\begin{align}
\widehat{x}_{k-1}&=\sum_{i=1}^{N}\widetilde{w}_{k-1}^{(i)}\widehat{x}_{k-1}(i),  \\
P_{k-1}&= \sum_{i=1}^{N}\widetilde{w}_{k-1}^{(i)} \big\{P_{k-1}(i)+\big(\widehat{x}_{k-1}-\widehat{x}_{k-1}(i)\big) \big(\cdot\big)^{T} \big\},
\end{align}	
\end{subequations}
where $(A)(\cdot)^{T}$ is a shorthand for  $AA^{T}$. Now having observed $y_k$, we would like to propagate the joint posterior in \eqref{factor_jt_pos} to time $k$, i.e.,
\begin{align}
p(x_{k-1},\lambda_{1:k-1}|y_{1:k-1}) \ {\stackrel{y_k}{\longrightarrow}}\ p(x_{k},\lambda_{1:k}|y_{1:k}). 
\end{align}
This can be achieved in the following steps ((1)-(4)) as described below:\\
%
%
%
\subsubsection{\textbf{PF prediction step}}
Generate $N$ new samples $\lambda_{k}^{(i)}$ from appropriately chosen proposal  $\pi(\lambda_{k}|\lambda_{1:k-1}^{(i)},y_{1:k})$. Then set $\lambda_{1:k}^{(i)} = \{\lambda_{1:k-1}^{(i)}, \lambda_{k}^{(i)}\}$, for $i=1,\cdots,N$, representing the particle trajectories up to time $k$. For proposal selection, see Remark \ref{remark_proposal} below.

\subsubsection{\textbf{KF prediction step}}
Since the prior $p(x_{k-1}|\lambda_{1:k-1}^{(i)},y_{1:k-1})$ for this step is Gaussian (given by \eqref{KF_output}) and the noises are  also Gaussian given $\lambda_{k}^{(i)}$, (see \eqref{cg_noises}), the dynamic system is now linear-Gaussian conditioned on $\lambda_{1:k}^{(i)}$. So, using KF, the predictive distribution 
can be obtained analytically, which is also Gaussian and given as $p(x_{k}|\lambda_{1:k}^{(i)},y_{1:k-1}) = \mathcal N\big(\widehat{x}_{k|k-1}(i), P_{k|k-1}(i)\big)$, where
\begin{eqnarray}
\widehat{x}_{k|k-1}(i)&=& A_k \ \widehat{x}_{k-1}(i) + B_k \ \mu^{w}_{k}(i) \label{KFpredm} \\
P_{k|k-1}(i)&=& A_k  \ P_{k-1}(i) \ A_k^{T}+ B_k \ Q_{k}(i)\ B_k^{T}\label{KFpredv}.
\end{eqnarray}
\subsubsection{\textbf{KF update step}}
Suppose we have now the new observation $y_k$. We can now update to the posterior distribution $p(x_{k}|\lambda_{1:k}^{(i)},y_{1:k})$, which is also Gaussian due to the CLGM, denoted as 
\begin{align}
p(x_{k}|\lambda_{1:k}^{(i)},y_{1:k})= \mathcal N\big(\widehat{x}_{k}(i), P_{k}(i)\big).	
\end{align}
The parameters $\widehat{x}_{k}(i)$, $P_{k}(i)$ can be obtained from the KF recursively, using the following steps:
\begin{eqnarray}
\widehat{x}_{k}(i) &=& \widehat{x}_{k|k-1}(i)+\nonumber \\
& & K_k(i)\big\{y_k -C_k\ \widehat{x}_{k|k-1}(i) -\mu^{e}_{k}(i)  \big\}\label{KFupm}\\
P_{k}(i) &=& P_{k|k-1}(i) - K_k(i)\ C_{k}^{T}\ P_{k|k-1}(i),
\end{eqnarray}
where
\begin{eqnarray}
K_k(i) &=& P_{k|k-1}(i)\ C_{k}^{T} \ S_k^{-1}(i)\\
S_k(i) &=& C_k \ P_{k|k-1}(i)\ C_{k}^{T}+ R_k(i)\label{KFups}. 
\end{eqnarray}
Moreover, the marginal likelihood is also obtained in closed form, 
which is also Gaussian and is given by
\begin{align}
p(y_{k}|\lambda_{1:k}^{(i)},y_{1:k-1})= \mathcal N\big(\mu^{L}_{k}(i),\ \Sigma^{L}_{k}(i)\big)\label{KFLkd},	
\end{align}
where
\begin{subequations}\label{KFLkd_param}
\begin{eqnarray}
\mu^{L}_{k}(i) &=& C_k \ \widehat{x}_{k|k-1}(i) + \mu^{e}_{k}(i)\\
\Sigma^{L}_{k}(i) &=& S_k(i).
\end{eqnarray}
\end{subequations}
\subsubsection{\textbf{PF update step}}
Now given the observation $y_k$ and the particles $\{\lambda_{1:k}^{(i)}\}_{i=1}^{N}$, we need to update to the posterior 
\begin{align} \label{PF_lambda_rep}
p(\lambda_{1:k}|y_{1:k})
  = \sum_{i=1}^{N}\widetilde{w}_{k}^{(i)}\delta(\lambda_{1:k}-\lambda_{1:k}^{(i)}); \ \widetilde{w}_{k}^{(i)} \ge 0,
\end{align}
with $\sum_{i=1}^{N}\widetilde{w}_{k}^{(i)}=1$. 
The corresponding weight $\widetilde{w}_{k}^{(i)}$ can be obtained (using \eqref{PF_gen_recursion}) as
\begin{subequations}	
\begin{align}
   w_{k}^{(i)}
  &\propto  
  \frac{p(y_{k}|\lambda_{1:k}^{(i)},y_{1:k-1})p(\lambda_{k}^{(i)}|\lambda_{1:k-1}^{(i)},y_{1:k-1})}
  {\pi(\lambda_{k}^{(i)}|\cdot)}\widetilde{w}_{k-1}^{(i)}, \label{pathspace_rec}\\
  \widetilde{w}_{k}^{(i)}&
 = \frac{{w}_{k}^{(i)}}{\sum_{j=1}^{N}{w}_{k}^{(j)}}.
\end{align}	
\end{subequations}
The density $p(y_{k}|\lambda_{1:k}^{(i)},y_{1:k-1})$ is already obtained in \eqref{KFLkd}. For now we assume that $p(\lambda_{k}^{(i)}|\lambda_{1:k-1}^{(i)},y_{1:k-1})$ is specified, which we discuss further in \ref{dyn_selection}.

This completes one propagation cycle. To propagate to the next cycle, we first resample the particles $\lambda_{1:k}^{(i)}$ (and also the associated $\big\{\widehat{x}_{k}(i),\ {P}_{k}(i)\big\}$) whenever necessary, before following again the steps (1)-(4).

\rem so long we can directly generate samples from $w_k$ and $e_k$, we can in principle, target the same inference task e.g., using a PF. However, when the state space is high dimensional, PF is well known to be inefficient. On the other hand, in our RBPF framework, PF targets only low dimensional state (auxiliary mixing vector, $\lambda_k$); conditionally the remaining state vector are treated using 
 KF. As a result, RBPF can scale well with the dimensions here.

%
%
%
%
\subsection{Specification of dynamics for $\lambda_{k}$}\label{dyn_selection}
As PF is essentially designed for the dynamically evolving model, we need to specify a proper  transition density $p(\lambda_{k}|\lambda_{1:k-1},y_{1:k-1})$, describing the time evolution of the path space $\lambda_{1:k}$. However this transition density is typically unknown and the only information we have is that $\lambda_{k}$ is distributed according to a (known) stationary density (say, $p^{\ast}$). We note that earlier in the context of symmetric $\alpha$-stable noise,  \cite{LombardiSJG06,Jordi11} treated $\lambda_{k}$ as unknown parameter with $p(\lambda_{k}|\lambda_{1:k-1},y_{1:k-1})=p(\lambda_{k}) \equiv p^{\ast}$ (i.e., $\lambda_{k}$  are generated independently over $k$). In the PF context, this is completely arbitrary and fundamentally weak, as path space based recursion (see e.g., \eqref{pathspace_rec}) requires a transition kernel, which links any newly generated particle to its ancestral lineage. To our knowledge, a provable state transition density in this context is not addressed adequately in the literature. 

In our approach, we specify the transition density through a Markov kernel such that the samples generated from this kernel constitute a Markov chain with the invariant distribution as marginal; since this invariant distribution is already known, we initialize the chain with this invariant distribution. In doing so, we follow \cite{Pittwalker05}, where the authors construct an AR($1$) process using the \textit{auxiliary latent variables}, such that the resulting time series is stationary with known marginals. Moreover, as shown in the original article, the time series model follows simple auto-correlation function, which decays exponentially. This property is in fact  very beneficial as it can limit the dependency of the sufficient statistics of $p(x_{k}|\lambda_{1:k},y_{1:k})$ on the particle path space. 

As shown in \cite{Pittwalker05}, based on the known invariant marginal, one can construct such a transition kernel easily, when $p(\lambda_{k})$ belongs to infinitely divisible convolution-closed exponential family and a family of density functions that includes among others, (inverse) gamma, normal and beta densities.
To keep the description simple, rather than going into the further details, we describe below one example to illustrate the idea. For more details, the reader may refer to \cite{Pittwalker05}.

\subsubsection{Example} note from \ref{sec:vg}, if $X$ is distributed according to a multivariate variance gamma, then the marginal $p(\lambda)$ is following an inverse gamma given by \eqref{VG-hier}-\eqref{IGdensity}. Suppose we are interested in constructing $p(\lambda_{k}|\lambda_{k-1})$. This can be implicitly obtained using the following relation\cite{Pittwalker05} 
\begin{align}
	\lambda_{k} = \frac{\nu/2 + U_k\ \lambda_{k-1}}{V_k},
\end{align}
 where $U_k \sim Ga(\alpha,1)$ and $V_k \sim Ga(\nu/2+\alpha,1)$, which are independent and also independent of each other. We also have
\begin{align}
	E(\lambda_{k}|\lambda_{k-1})= \mu(1-\rho) + \rho \lambda_{k-1}, 
\end{align}
where $\rho = \frac{2\alpha}{\nu+2\alpha-2}$ is the autocorrelation function\footnote{The autocorrelation function is restricted to be positive (i.e., $\rho \in (0,1)$) for such construction.} and $\mu = \frac{\nu}{\nu-2}$ is the mean of the marginal distribution. Note that $E(\lambda_{k}|\lambda_{k-1})$ exists for $\nu>2$ and the autocorrelation function of $\lambda_{k}$  is $\rho(\tau) = \rho^{\tau}$. Thus, the transition kernel can be specified by a first order AR model with inverse-gamma marginals. The unknown parameter $\alpha$ (associated with $U_k$ and $V_k$) is specified here through the selected value of $\rho$.

\rem\label{remark_proposal}The optimal proposal \cite{Saha_SC} given by $\pi(\lambda_{k}|\lambda_{1:k-1},y_{1:k})= \pi(\lambda_{k}|\lambda_{k-1},y_{k})$ is often unavailable in practice. One simple  alternative is to implement a \textit{bootstrap} PF with transition kernel $\pi(\lambda_{k}|\lambda_{k-1})$ as the proposal.
%
%
\subsection{Algorithmic summary }\label{alg:known_noise}

A \textit{bootstrap} implementation of the proposed RBPF for the LDS with hierarchical Gaussian noises are summarized in Algorithm~\ref{alg:rbpf}.
\begin{algorithm}
  \caption{RBPF for LDS}\label{alg:rbpf}
	Set HGM for process and/or measurement noises\\
	 - $p(\lambda_k)$ is set accordingly\\
	For each particle $i=1,\dots,N$ do \\
  \underline{Initialization:}
    \begin{itemize}[noitemsep,topsep=0pt,parsep=0pt,partopsep=0pt]
  \item set $p(x_{0}^{(i)})$ for each KF
  \end{itemize}

  \underline{Iterations:}\\
	For $k=1,2,\ldots$ do
	\begin{enumerate}
	\item PF prediction step
    \begin{itemize}[noitemsep,topsep=0pt,parsep=0pt,partopsep=0pt]
     \item sample $\lambda_{k}^{(i)}$ according to section \ref{dyn_selection}
	      \begin{itemize}[noitemsep,topsep=0pt,parsep=0pt,partopsep=0pt]
		    \item if $k=1$, $\lambda_{k}^{(i)}\sim p(\lambda_{k})$
	      \item else, $\lambda_{k}^{(i)}\sim p(\lambda_{k}|\lambda_{k-1}^{(i)})$
       \end{itemize}
	
	   \item set $\lambda_{1:k}^{(i)} \triangleq (\lambda_{1:k-1}^{(i)},
      \lambda_{k}^{(i)})$
	\end{itemize}		
\item KF prediction step
	 \begin{itemize}[noitemsep,topsep=0pt,parsep=0pt,partopsep=0pt]		
	\item set $\mu^{w}_{k}(i),\mu^{e}_{k}(i), Q_{k}(i), R_{k}(i)$ (Eq. \eqref{cg_noises})
	\item compute $\widehat{x}_{k|k-1}(i)$ and $P_{k|k-1}(i)$ (Eq. \eqref{KFpredm}-\eqref{KFpredv}) 
  	\end{itemize}
\item KF update step	
	 \begin{itemize}[noitemsep,topsep=0pt,parsep=0pt,partopsep=0pt]		
	\item compute $\widehat{x}_{k}(i)$ and $P_{k}(i)$ (Eq. \eqref{KFupm}-\eqref{KFups})
	\item compute $p(y_{k}|\lambda_{1:k}^{(i)},y_{1:k-1})$ (Eq. \eqref{KFLkd}-\eqref{KFLkd_param})
	\end{itemize}
\item PF update step	
  \begin{itemize}[noitemsep,topsep=0pt,parsep=0pt,partopsep=0pt]		
	\item weight update 
	   \begin{itemize}
	  \item $w_{k}^{(i)}\propto p(y_{k}|\lambda_{1:k}^{(i)},y_{1:k-1})$
		\item normalize weights
		\end{itemize} 
	\item resample the particles
	\begin{itemize}
	  \item let resampled particle indices be $j_{i}\in \{1,\cdots,N\}$
		\item set $\widehat{x}_{k}(i)=\widehat{x}_{k}(j_{i})$ and $P_{k}(i)=P_{k}(j_{i})$
	
		\item set $\lambda_{1:k}^{(i)}$ = $\lambda_{1:k}^{(j_{i})}$
		\item  move to the next step
		\end{itemize} 
	\end{itemize}

	\end{enumerate} 
\end{algorithm}

\subsection{Estimation of the likelihood function}
\noindent
The likelihood function is the joint density of the observation $y_{1},\cdots,y_{T},$ which can be decomposed as
\begin{align}
	p(y_{1},\cdots,y_{T})= \prod_{k=1}^{T} p(y_{k}|y_{1:k-1}),
\end{align}
where for $k=1$, $p(y_{k}|y_{1:k-1})$ is interpreted as $p(y_1|\varnothing)\equiv p(y_1)$. The density $p(y_{k}|y_{1:k-1})$ can be obtained by marginalizing $\lambda_{1:k}$ in \eqref{KFLkd} using \eqref{PF_lambda_rep} as
\begin{align}
p(y_{k}|y_{1:k-1})\approx \sum_{i=1}^{N}\widetilde{w}_{k}^{(i)} \mathcal N\big(\mu^{L}_{k}(i),\ \Sigma^{L}_{k}(i)\big).	
\end{align}
%
%
%

\subsection{p-step ahead prediction}
\noindent
Suppose at time step $k$, we have the observations $\{y_{1},\cdots,y_{k}\}$ and the corresponding filter density $p(x_{k}|y_{1:k})$. The p-step ahead prediction of the state can now be obtained as follows:
\begin{eqnarray}
p(x_{k+p}|y_{1:k}) & = & \int p(x_{k+p}|y_{1:k}, \lambda_{1:k}) \ p(\lambda_{1:k}|y_{1:k}) d\lambda_{1:k} \nonumber \\
&\approx&\sum_{i=1}^{N}{w}_{k}^{(i)} p(x_{k+p}|y_{1:k}, \lambda_{1:k}^{(i)}).
\end{eqnarray}

Now $p(x_{k+p}|y_{1:k}, \lambda_{1:k}^{(i)})$ can be obtained using p-step ahead KF prediction as
\begin{align}
	p(x_{k+p}|y_{1:k}, \lambda_{1:k}^{(i)})= \mathcal N(\mu_{k+p|k}^{(i)},\Sigma_{k+p|k}^{(i)}),
\end{align}
with
\begin{equation}
\mu_{k+p|k}^{(i)} = \Phi_{k+1+p,k+1}\ \widehat{x}_k(i)\ + 
  \quad \sum_{j=1}^{p}\Phi_{k+1+p,k+1+j}\ B_{k+j}\ \mu_{k+j}^{w}(i),
\end{equation}
\begin{multline}
\Sigma_{k+p|k}^{(i)}= \Phi_{k+1+p,k+1}\ P_k(i)\ (\Phi_{k+1+p,k+1})^{T}\ +  \\
  \sum_{j=1}^{p}\Phi_{k+1+p,k+1+j}\ B_{k+j} Q_{k+j}(i)(B_{k+j})^{T}(\Phi_{k+1+p,k+1+j})^{T},
\end{multline}
where we use the notation $\Phi_{k,l}= A_{k-1}A_{k-2}\cdots A_{l}\ \ (k>l)$ and $\Phi_{k,k}=I$\cite{Anderson_79}.
%
%
%
%
%
%
%

%
%

\section{Robust noise adaptive filtering for LDS}\label{rob_noise_adap}
\noindent
Although, the distribution of the noise is assumed to be stationary and known in the previous section, it is important to observe that the proposed algorithm is already doing a noise adaptive filtering. To see this, first note from \eqref{PF_marginal} that we are also estimating  $\lambda_{k}$ sequentially over time. Now moving from time $(k-1)$ to $k$, when we generate ${N}$ new samples $\{\lambda_{k}^{(i)}\}_{i=1}^{N}$, the empirical distribution associated with $p(\lambda_{1:k})$ is given by the particle cloud $\big\{ \lambda_{1:k}^{(i)},\ (1/N) \big\}_{i=1}^{N}$. Since the noise in \eqref{cg_noises} at time $k$ is assumed to be hierarchically Gaussian,  we can now represent the density for the noise (say $v_k$) as 
\begin{equation}\label{HGM_MC_rep}
	p(v_k) \approx \frac{1}{N} \sum_{i=1}^{N} \calN \bigg(v_k|\mu(\lambda_{1:k}^{(i)}), \Sigma(\lambda_{1:k}^{(i)})\bigg),
\end{equation}
which is an equally weighted finite (random) Gaussian mixture (with component weight $=1/N$). This serves as our prior for the (unknown) noise. When the new measurement $y_k$ arrives, we update to noise posterior using the Bayes rule. The posterior now consists of the same random mixture components, but the component weights are adjusted according to the observation likelihood (i.e. $p(y_{k}|\lambda_{1:k}^{(i)},y_{1:k-1})$). This construction ensures that we have a non-Gaussian noise adaptive filter.  However, the stability of this filter is still  sensitive to any potential large noise that cannot be accounted by the selected prior. In practice, the robustification for this noise adaptive filter is done through a flat prior selection. This ensures that the resulting filter is robust to outliers, yet at the same time due to its noise adaptive behavior, the performance does not degrade when the outliers are absent. 

%
%
%
%
%
%
%

\section{Numerical studies}\label{num_experiment}
\noindent
We start here with the inference problem when the measurement noise is skewed but its distribution is known. This is illustrated for a stochastic volatility problem (finance). Next we consider the cases where noise parameters are unknown and time varying. Here we test the proposed robust noise adaptive filter on a time series problem for the following cases: (a) unknown measurement noise and (b) both process and measurement noises are unknown. Finally we consider a maneuvering target tracking example and compare the proposed approach with the interacting multiple model (IMM) algorithm. 

\subsection{LDS with known stationary non-Gaussian noise }
\begin{figure}[ptbh]
  \centering
 	\includegraphics[width=4.0in]{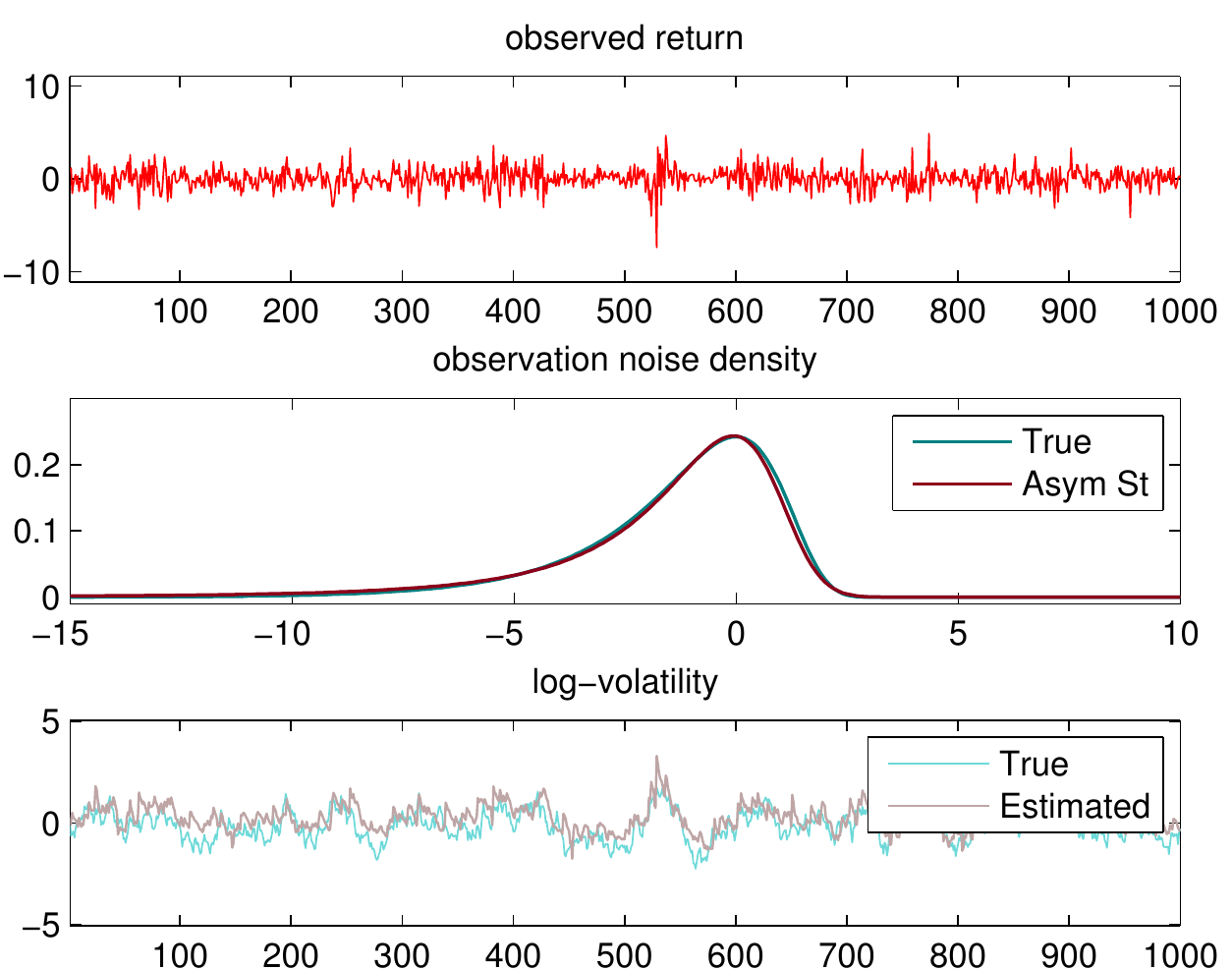}
  \caption{Stochastic volatility example: (top) simulated return data $y_k$, (middle) observation noise density and its GH skew Student's \textit{t} approximation, (bottom) the true log-volatility and one realization of the filter mean estimates.}
  \label{fig:Skewd_GH_noise}
\end{figure}

We consider the online filtering problem for the volatility example defined by \eqref{LSV:dyn}-\eqref{LSV:obs}.  This example illustrates the strength of the proposed framework in terms of treating a skewed noise (although we note that  PF is more appropriate for this scalar problem when computational cost is considered).  
As is evident from Figure (\ref{fig:Skewd_noise}), the observation noise is highly skewed. In the literature, this has been approximated e.g., by a mixture of seven Gaussian components by equating the first four moments \cite{Kim98}. Here we approximate this noise density using a GH skew Student's \textit{t} distribution with $\mu=1.75$, $\beta=-2.3$, $\Sigma=1$ and $\nu=5.8$. Following \cite {Shephard96}, we select $\gamma_0=0$, $\gamma_1=0.9$ and $\sigma_{n}^{2}=0.1$ and using \eqref{SVmodel},  generate a (simulated) return data $y_k$ for $1000$ time steps. Next, using the simulated data, we estimate the filter distribution by Algorithm\ref{alg:rbpf}. We observed that the algorithm is handling well the sequential estimation task. The simulated data, the (approximated) observation noise density  and a typical realization for the filter mean estimates are plotted in Figure (\ref{fig:Skewd_GH_noise}). 
\subsection{Time varying and unknown noise parameters}
Here we start with the case of unknown measurement noise, which is then followed by the case, where both the process noises and measurement noises are unknown.
\subsubsection{Measurement noise is unknown}
\begin{figure}[ht]
\centering
\begin{minipage}[b]{0.45\linewidth}
\includegraphics[width=3.0in]{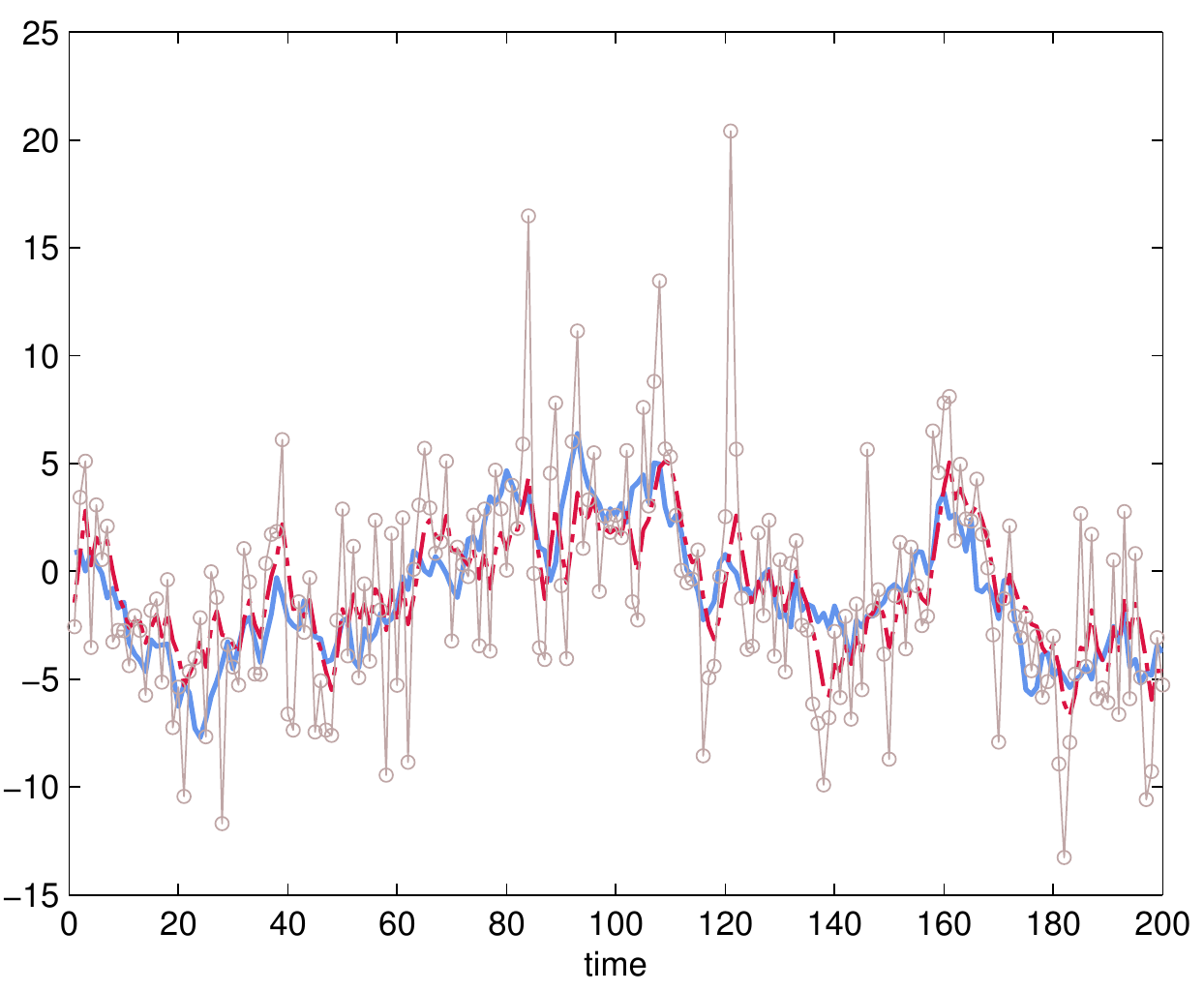}
\caption{When measurement noises containing sporadic outliers; true signal ('$-$'), measurements ('$-o$') and mean estimates of the signal ('$-.$').}
\label{fig:Toy_O_nonpers}
\end{minipage}
\quad
\begin{minipage}[b]{0.45\linewidth}
\includegraphics[width=3in]{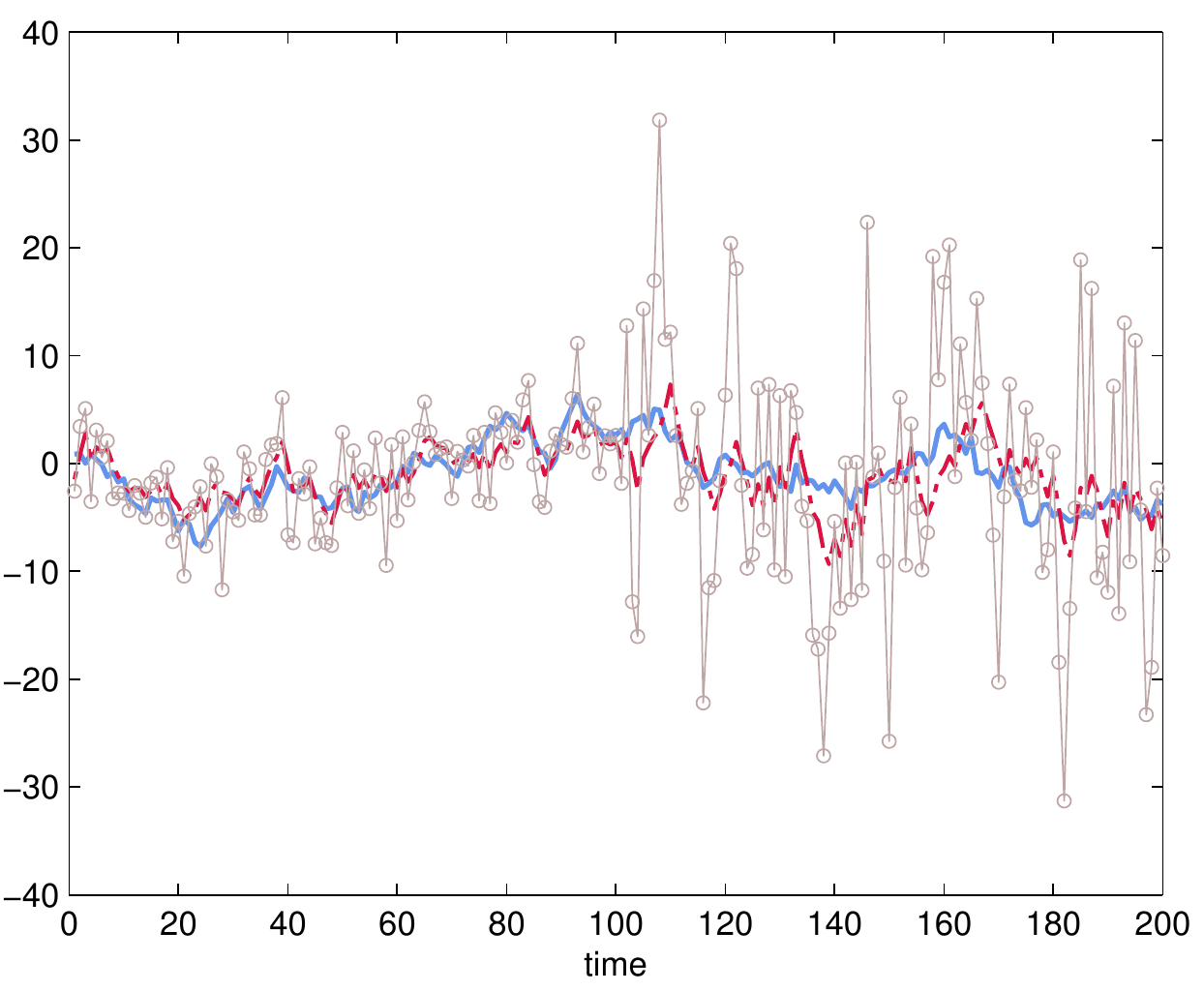}
\caption{When measurement noises containing persistent outliers; true signal ('$-$'), measurements ('$-o$') and mean estimates of the signal ('$-.$').}
\label{fig:Toy_O_pers}
\end{minipage}
\end{figure}

we consider the following time series problem given by
\begin{subequations}\label{TSprob}
\begin{align}
s_k &= 1.51 s_{k-1}-0.55 s_{k-2}+ w_k, \\ 
y_k &= s_k + e_k, 	
\end{align}
\end{subequations}
where the signal $s_k$ is evolving as a second order auto-regressive process, with $w_k \sim \mathcal{N}(0,1)$. The distribution for the measurement noise $e_k$ is assumed to be unknown. The simulated (synthetic) measurements $y_k$ are generated using the mixture distribution $e_k \sim 0.95\mathcal N(0,10)+ 0.05\mathcal N(0,100)$. For the robust noise adaptive filter, the prior for unknown $e_k$ is taken as a zero mean Student's \textit{t} with $\Sigma =10^4$ and $\nu=5$. A typical realization of the filter mean estimate is shown in Figure (\ref{fig:Toy_O_nonpers}).  
One can observe that the measurements contain outliers that are appearing sporadically (statistically independent from one another), but the filter is doing a really good job in estimating the hidden signals. 

But there may be situation where the outliers can be persistence (time correlated). For example, noise may temporarily switch to a different regime (with a higher covariance) due to sensor malfunction. This change in noise cannot be counted normally by a nominally specified noise model. We consider this scenario next. For the simulated observations, we assume  that the measurement noises for the first $100$ time steps is distributed as $e_k \sim \mathcal N(0,10)$. Then due to the sensor malfunctioning, the measurement noises for the next $100$ time steps is distributed as $e_k \sim \mathcal N(0,100)$. However, this knowledge is not available to the filter. Again the prior for unknown $e_k$ is taken to be a zero mean Student's \textit{t} with $\Sigma =10^4$ and $\nu=5$ and  a typical realization of the filter mean estimate is shown in Figure (\ref{fig:Toy_O_pers}). In this case as well, the proposed filter is doing a good job in estimating the signals.
%
%
%
\subsubsection{Both process and measurement noises are unknown}
\begin{sidewaysfigure}
  \centering
  \begin{subfigure}{0.45\textheight}
    \includegraphics[width=0.45\textheight,height=7cm]{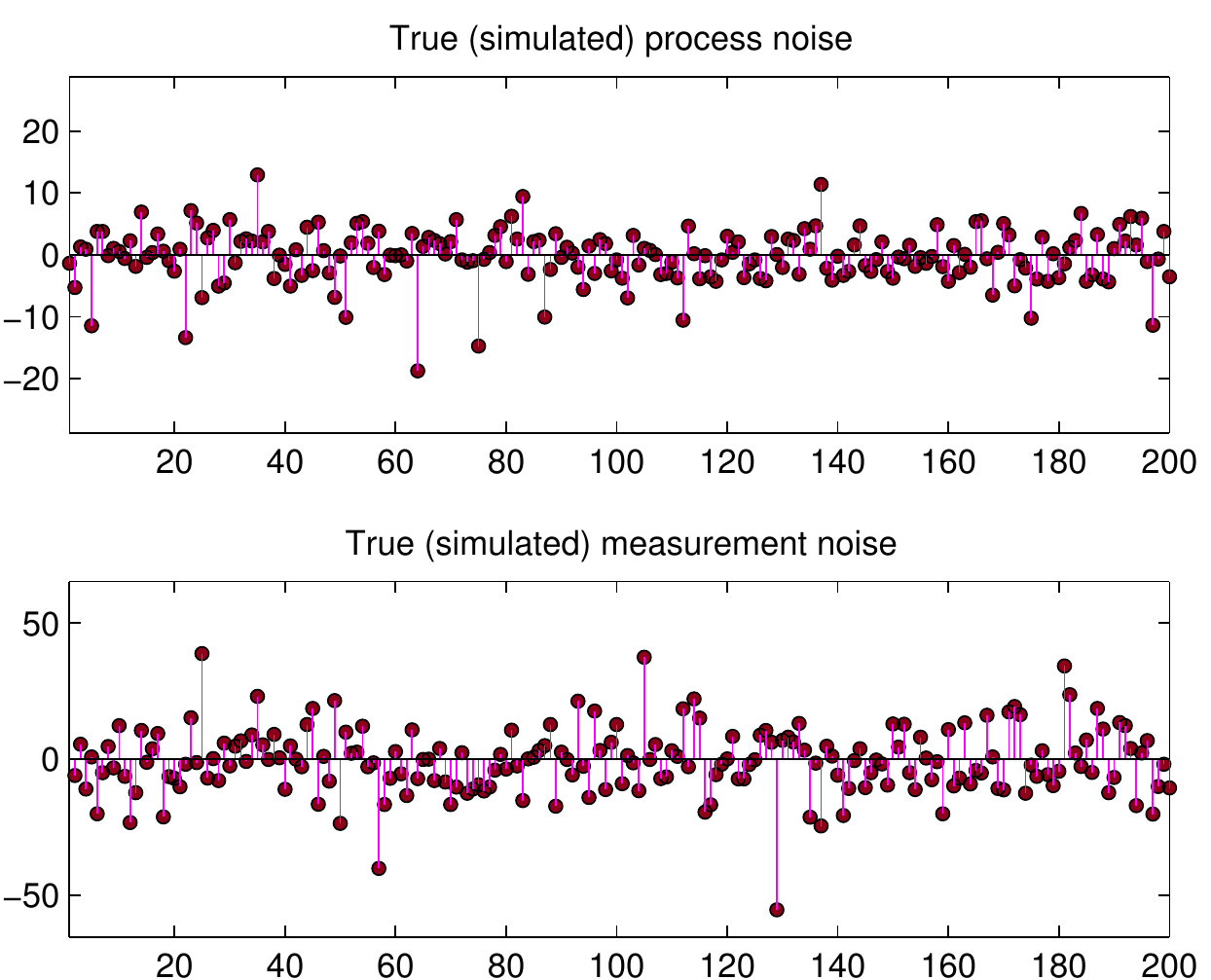}
    \caption{Sporadic outliers: outliers in both noises.}
  \label{fig:Toy_B_nonpers_noise}
  \end{subfigure}
	  \begin{subfigure}{0.45\textheight}
    \includegraphics[width=0.45\textheight,height=7cm]{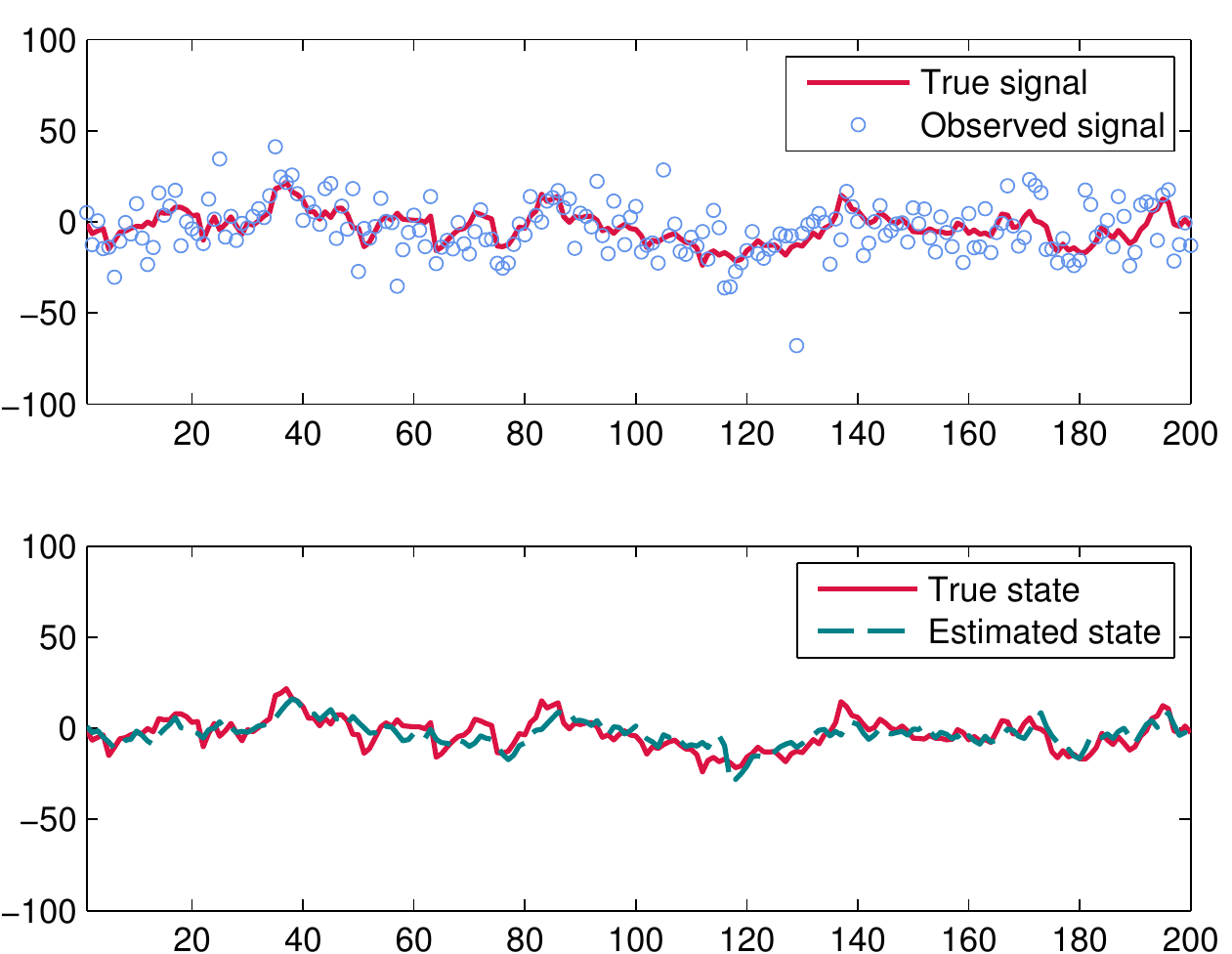}
    \caption{Sporadic outliers:(estimated) states and measurements.}
  \label{fig:Toy_B_nonpers_filt}
  \end{subfigure}
	  \begin{subfigure}{0.45\textheight}
    \includegraphics[width=0.45\textheight,height=7cm]{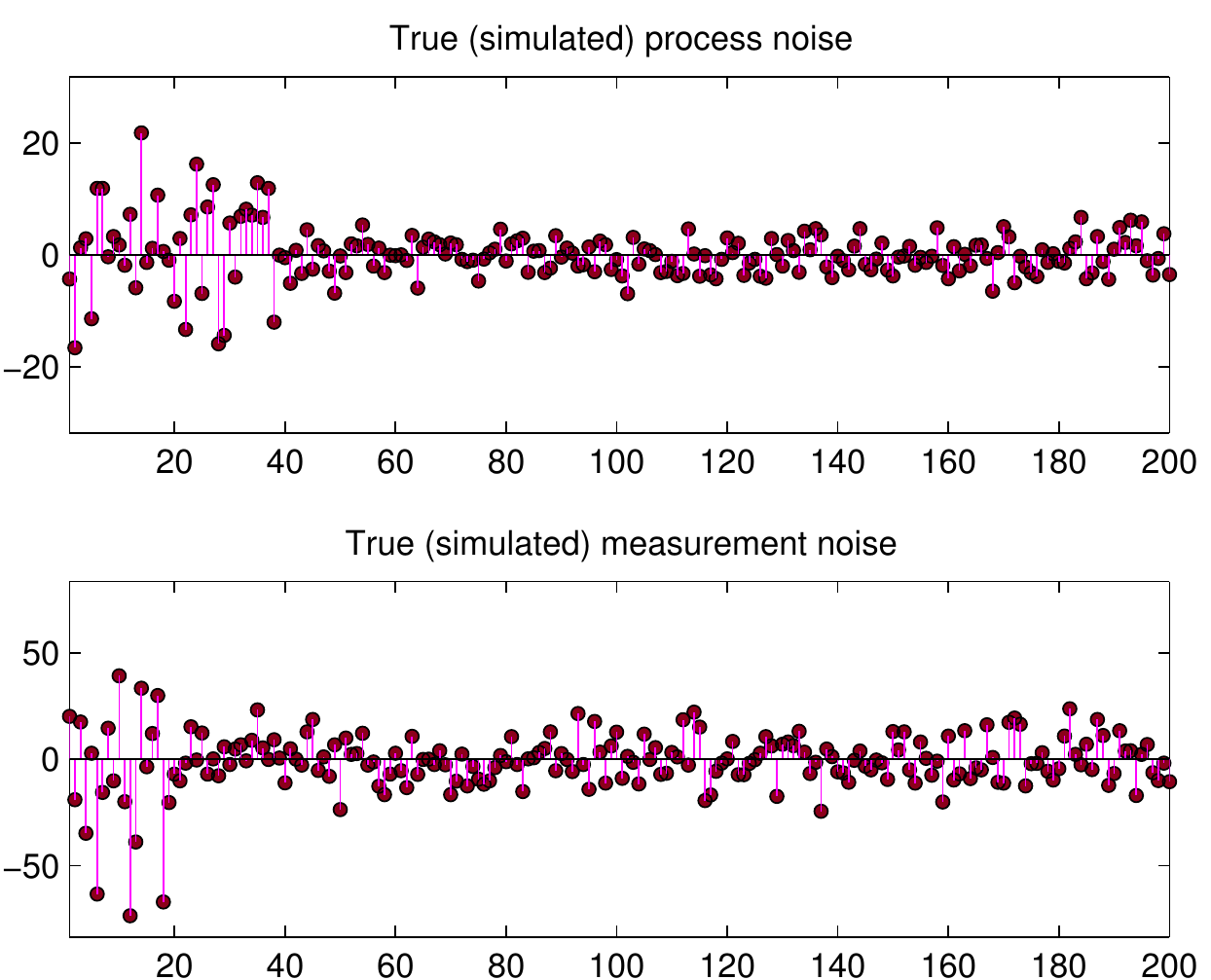}
    \caption{Persistent outliers: outliers in both noises.}
  \label{fig:Toy_B_pers_noise}
  \end{subfigure}
	\begin{subfigure}{0.45\textheight}
    \includegraphics[width=0.45\textheight,height=7cm]{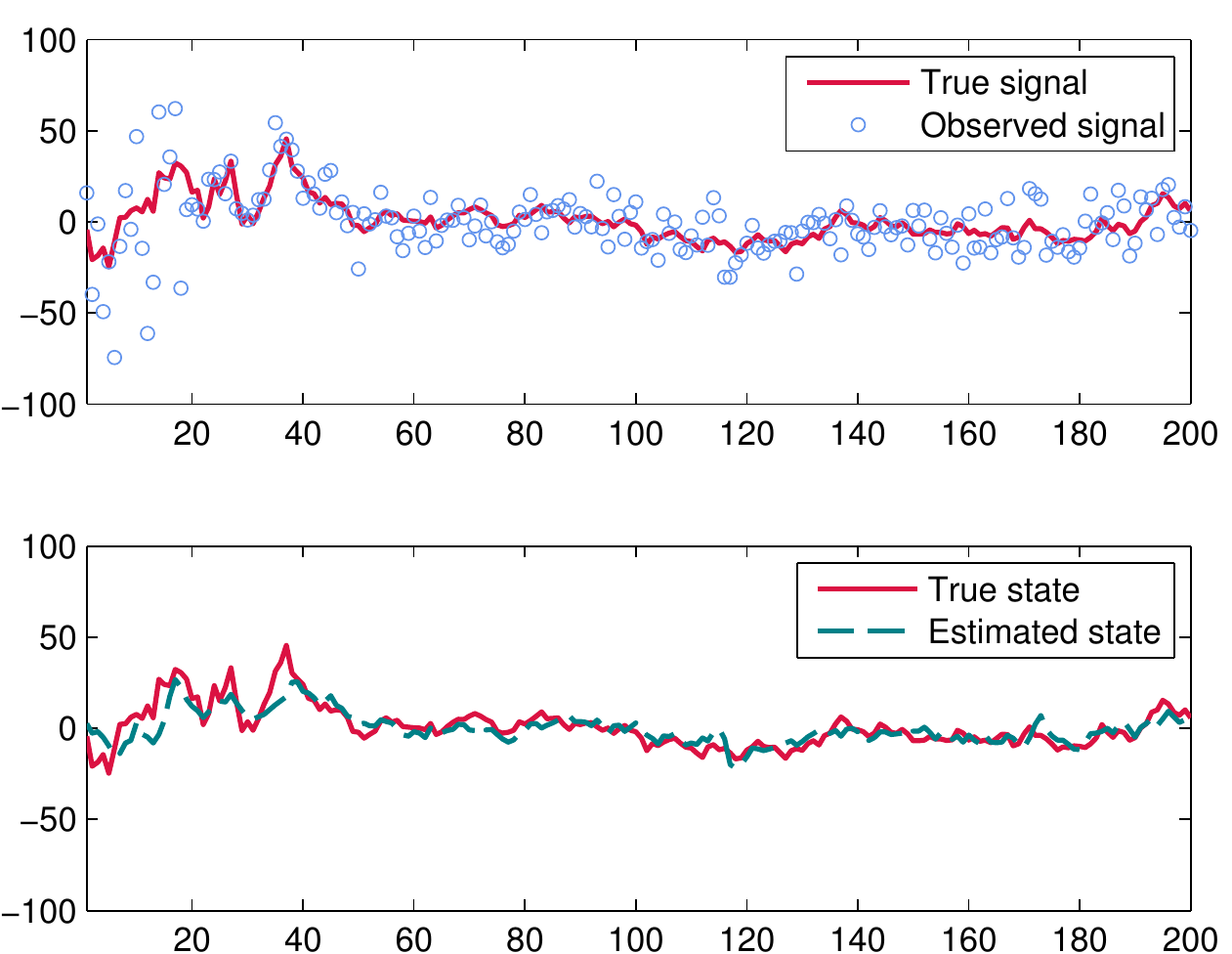}
    \caption{Persistent outliers:(estimated) states and measurements.}
  \label{fig:Toy_B_pers_filt}
  \end{subfigure}
  \caption{Filtering when both the process and the measurement noises contain outliers}
\end{sidewaysfigure}

for this problem, although the framework is  relatively simple (i.e., use HGM priors for both the noises as in \eqref{cg_noises}), we note that 
the inference problem can be difficult, at least for certain problems, due to identifiability issue. For example, if for any running KF (indexed by \textit{i}), there is any larger than the usual value of innovation (defined as $\big\{y_k -C_k\ \widehat{x}_{k|k-1}(i) -\mu^{e}_{k}(i)\big\}$ in \eqref{KFupm}), KF cannot immediately distinguish whether this is due to  any observation outlier or structural break in the process model. Nonetheless, filtering with unknown heavy tailed process and measurement noises have earlier been considered in e.g., \cite{Godsill96,Masreliez75,Roth_Lic} and so we include some numerical studies as well. We consider the LDS as in \eqref{TSprob}, where the true (but unknown) noises are distributed as $w_k \sim 0.8\mathcal N(0,10)+ 0.2\mathcal N(0,100)$ and $e_k \sim 0.9\mathcal N(0,100)+ 0.1\mathcal N(0,1000)$. For filtering, the state is estimated using $50$ particles and the noise priors as $p(w_k) = \mathcal{T}(0,10^4,5)$ and $p(e_k) = \mathcal{T}(0,10^5,5)$ respectively. The true (generated) noise realizations are shown in Figure (\ref{fig:Toy_B_nonpers_noise}) while the true and a typical estimated (mean) state along with measurements are shown in Figure (\ref{fig:Toy_B_nonpers_filt}). Next, we consider the case of persistent noise outliers with the same filter settings. Here the process noises for first $40$ step is generated with $\mathcal N(0,100)$ and the rest with $\mathcal N(0,10)$. Similarly the measurement noises for first $20$ step is generated with $\mathcal N(0,1000)$ and the rest with $\mathcal N(0,100)$. The true (generated) noise realizations for this case are shown in Figure (\ref{fig:Toy_B_pers_noise}). Corresponding true and a typical estimated (mean) state along with measurements are shown in Figure (\ref{fig:Toy_B_pers_filt}). For both cases, we observe that the filter is performing reasonably well. 
%
%
%
%
%
%
%
%
%
\subsection{Tracking a maneuvering target}
\begin{figure}[ptbh]
  \centering
 	\includegraphics[width=3in]{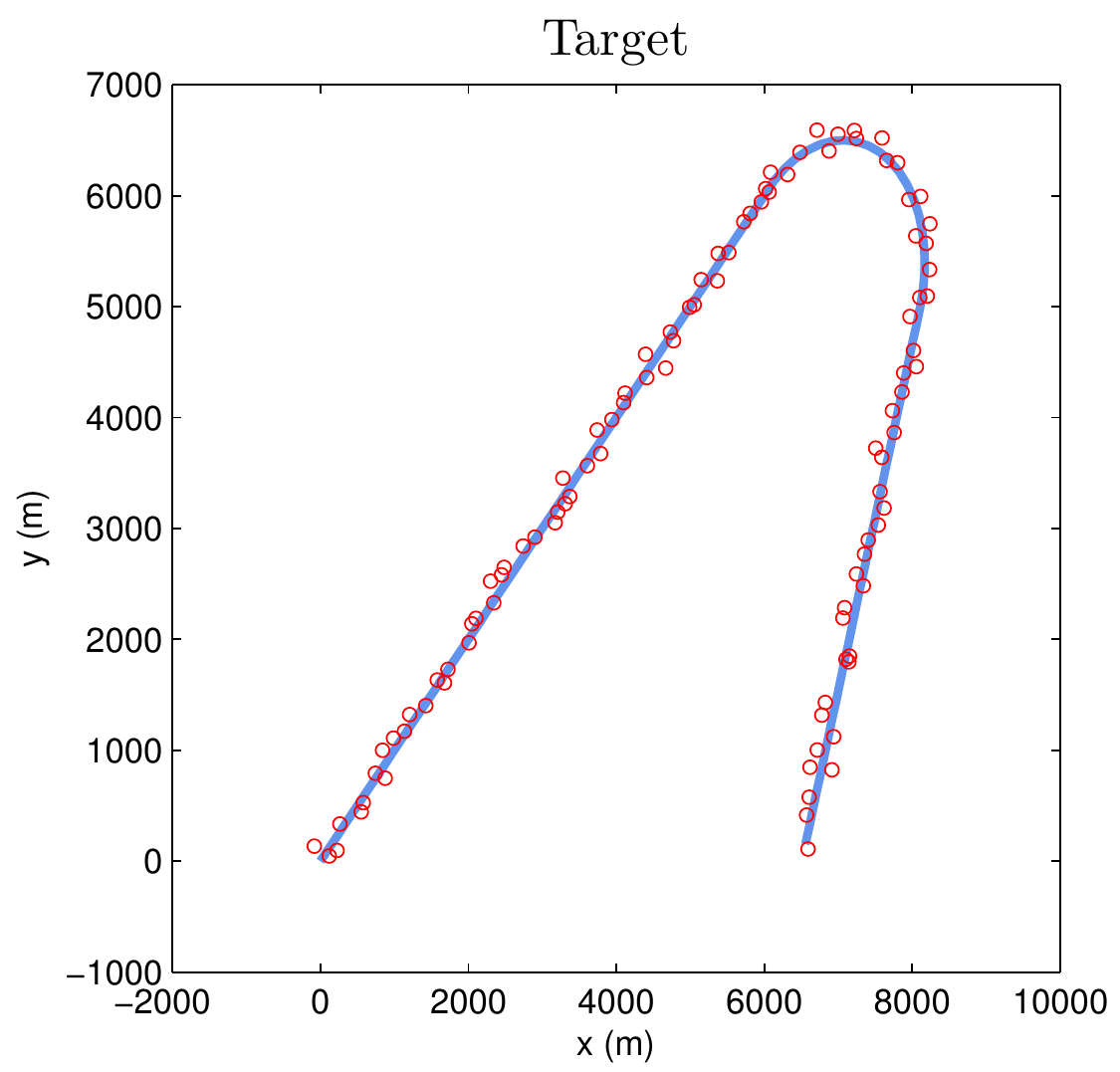}
  \caption{The true positions ('$-$') and the measured positions ('$o$') of the target (the target starts at (0,\ 0)).}
  \label{fig:true_track}
\end{figure}

\begin{sidewaysfigure}
    \centering
    \begin{subfigure}{0.4\textheight}
        \centering
        \includegraphics[width=0.4\textheight,height=7cm]{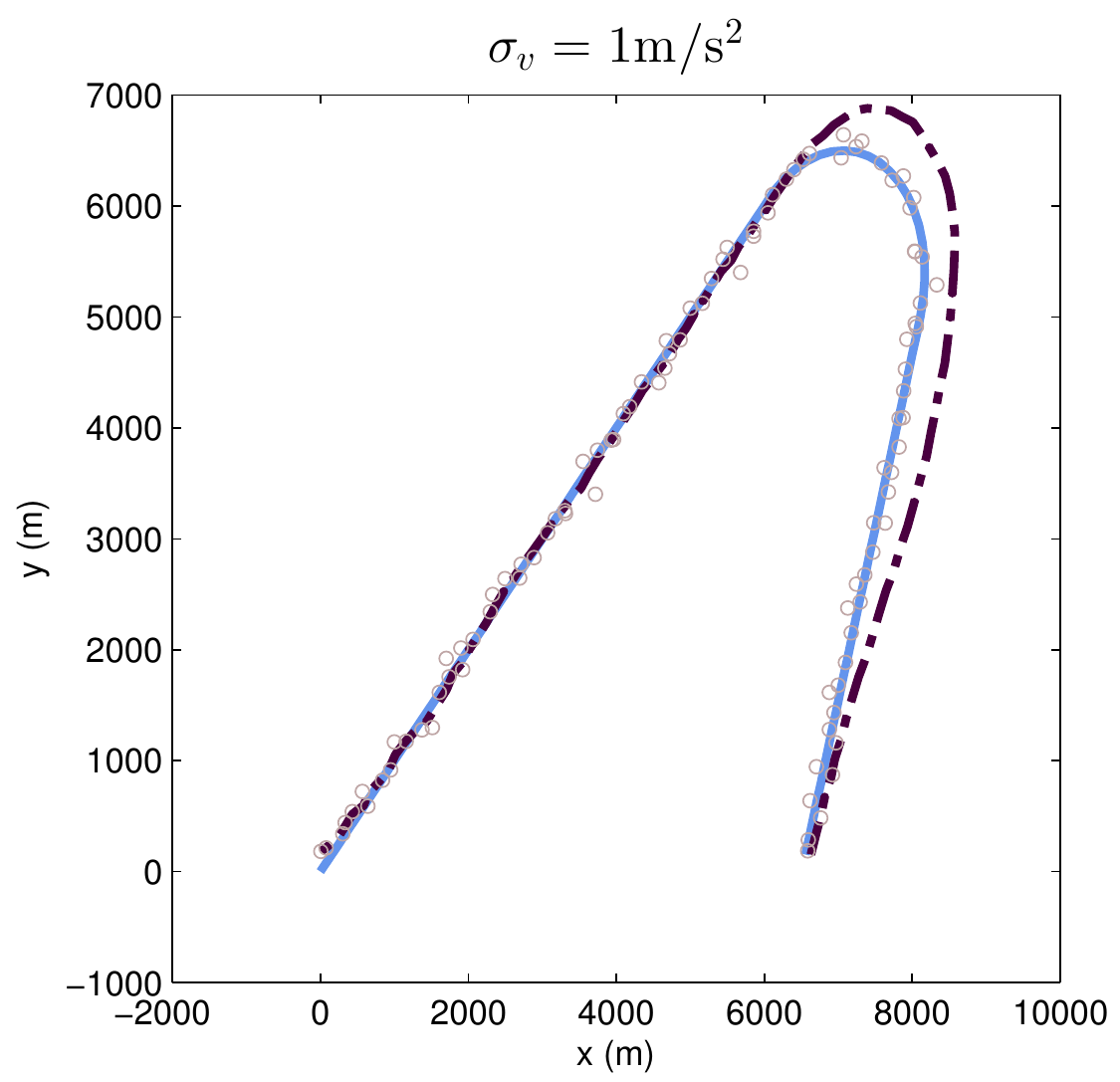}
        \caption{CV model: true track ('$-$'), estimated track ('$-.$'), measurements ('$o$').}
		\label{fig:MTT_smallproc}		
    \end{subfigure}%
		\begin{subfigure}{0.4\textheight}
        \centering
        \includegraphics[width=0.4\textheight,height=7cm]{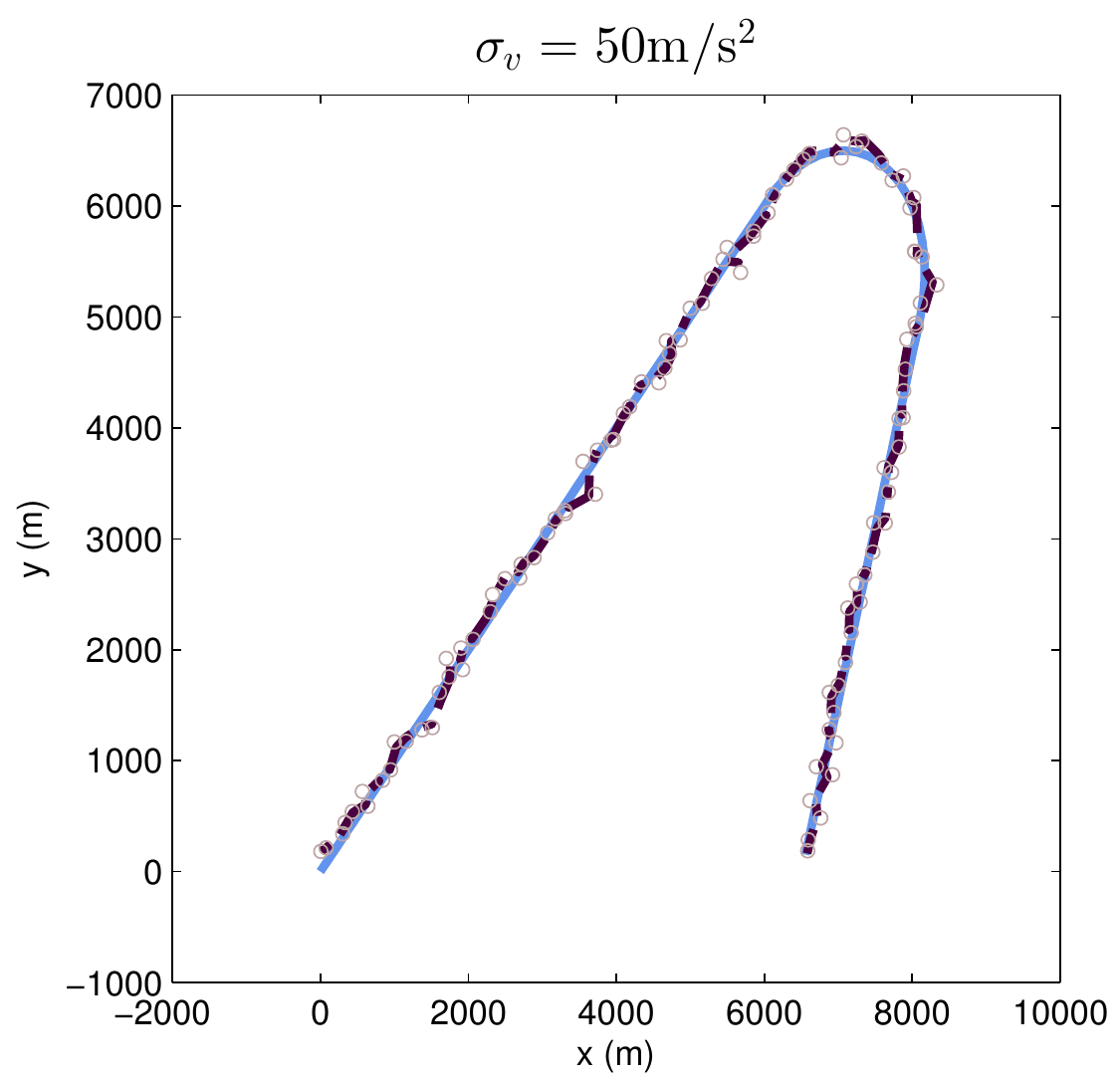}
        \caption{CV model: true track ('$-$'), estimated track ('$-.$'), measurements ('$o$').}
				\label{fig:MTT_bigproc}
    \end{subfigure}
		\begin{subfigure}{0.4\textheight}
        \centering
        \includegraphics[width=0.4\textheight,height=7cm]{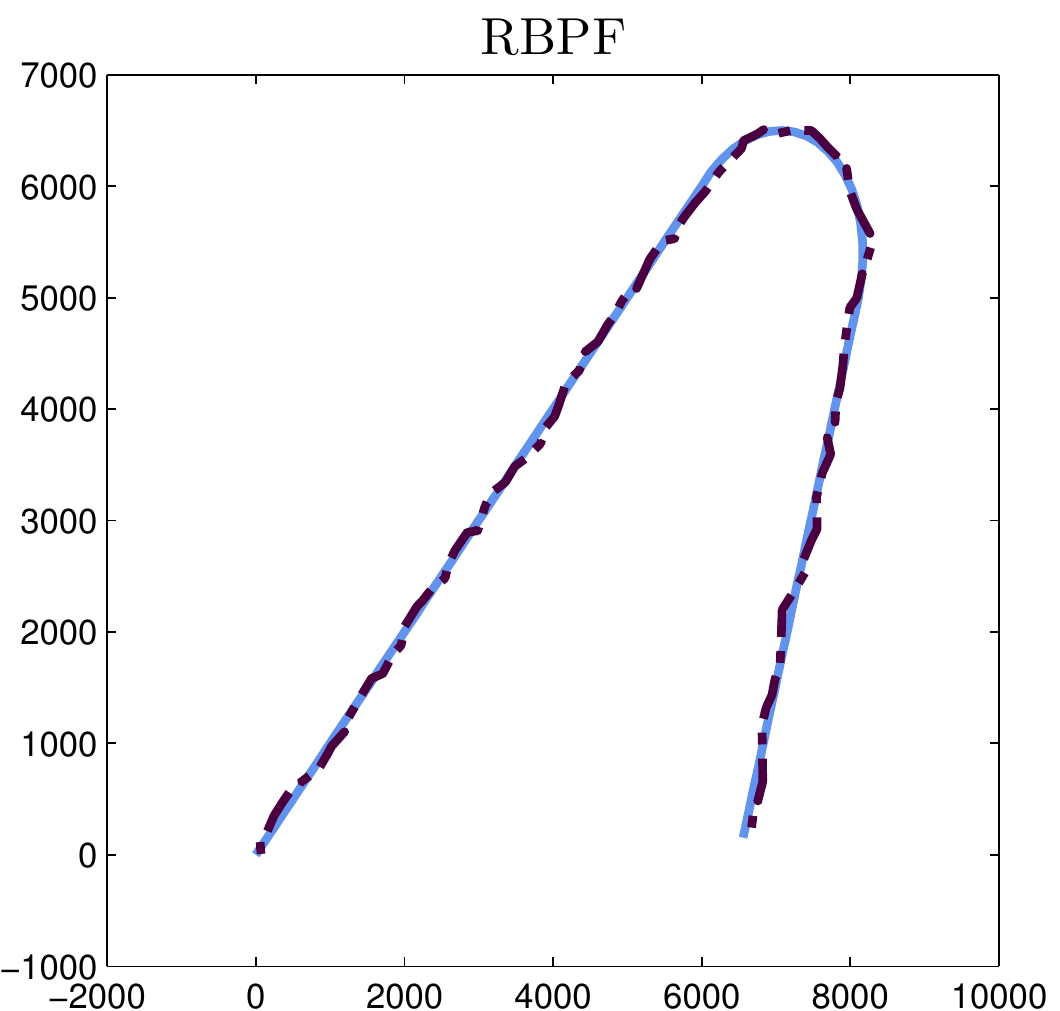}
        \caption{Robust noise adaptive model: true track ('$-$'), estimated track ('$-.$').}
				\label{fig:RBPFproc}
    \end{subfigure}
		\begin{subfigure}{0.4\textheight}
        \centering
        \includegraphics[width=0.4\textheight,height=7cm]{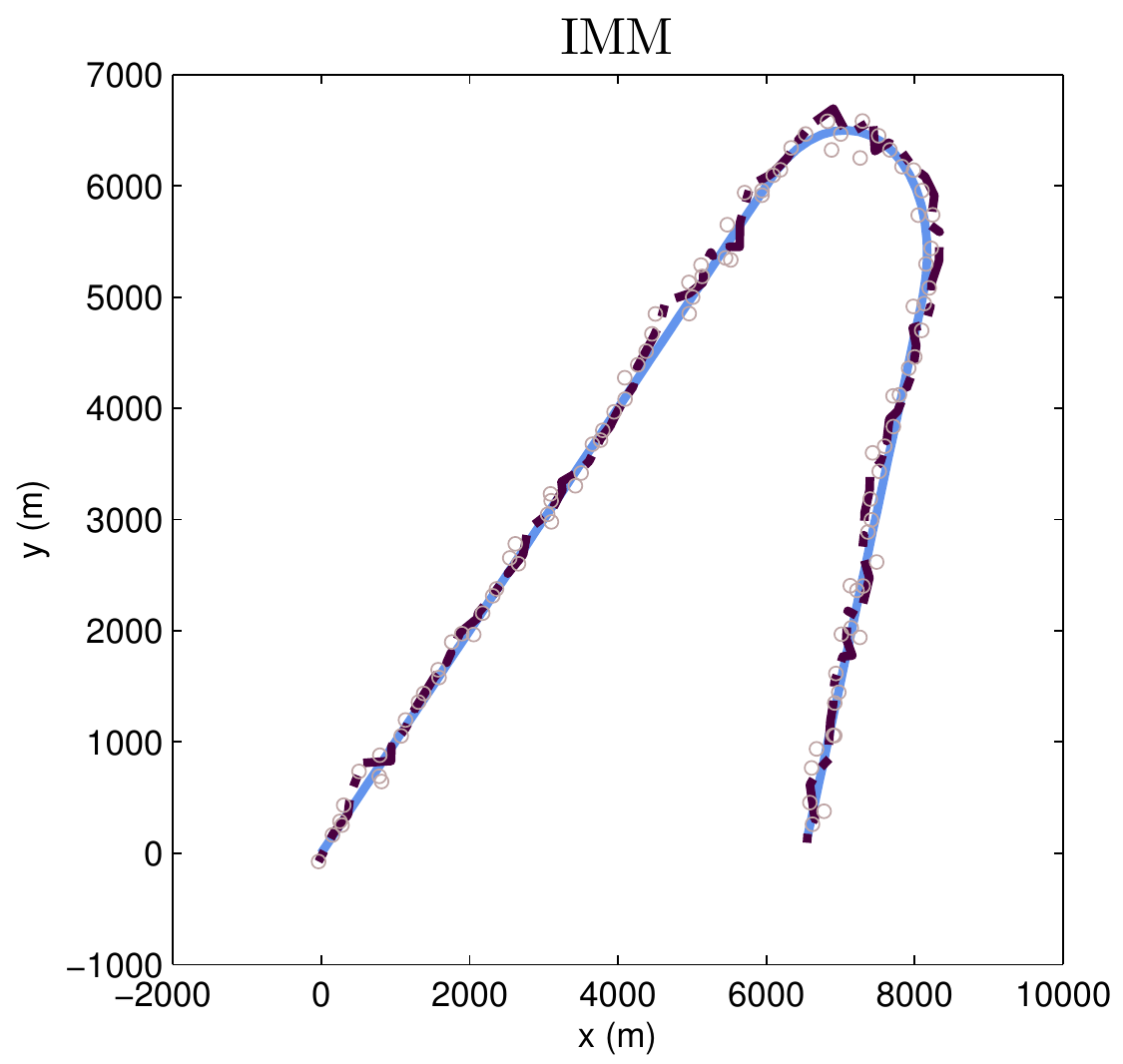}
        \caption{IMM: true track ('$-$'), estimated track ('$-.$'), measurements ('$o$').}
				\label{fig:IMM}
    \end{subfigure}
    \caption{Maneuvering target tracking example; true and estimated trajectory of the target}
\end{sidewaysfigure}

We consider a (simplified) problem of tracking a maneuvering target in two dimensional plane, where the state vector $(x_k)$ consists of the Cartesian position and velocity. The noise corrupted snapshots of the positions $(y_k)$ are available as the measurements. The target starts at $(0,0)$ and initially proceeds with a linear motion. This is followed by a (clockwise) coordinated turn and then again a linear motion. The true target trajectory and the noisy measurements are shown in Figure (\ref{fig:true_track}). We propose to track the target first, using a so called constant velocity (CV) model \cite{Risticbook04} given by
\begin{eqnarray}
x_{k+1}&=&\left[\begin{array}{cc}I_{2}&TI_{2}\\0&I_{2}\end{array}\right]x_k+\left[\begin{array}{c}\frac{T^{2}}{2} I_{2}\\T I_{2}\end{array}\right]v_{k} \\
y_k &=& \left[I_{2} \ \ 0\right]x_k + e_k,
\end{eqnarray}
where $T$ is the sampling time in second (we use $T=1$), $I_2$ is a two dimensional identity matrix, $v_k$ and $e_k$ are the process and the measurement noise respectively. We assume that $e_k$ is generated according to a known distribution, given by $e_k\sim \mathcal{N}(0,diag[80^{2}, \ 80^{2}])$, which is also used to generate the true measurements here. However due to multi-modality in the track behavior, an appropriate process noise is difficult to specify. This point is illustrated by estimating the track using two different process noise standard deviations, i.e., $\sigma_{v}=1$ and $\sigma_{v}=50$ (assuming $v_k$ to be Gaussian). The results are shown Figure (\ref{fig:MTT_smallproc}) and Figure (\ref{fig:MTT_bigproc}) respectively.   
We see that $\sigma_{v}=1$ is a better choice for the (initial) linear part of the trajectory, whereas $\sigma_{v}=50$ is more suitable during the maneuvering part. Thus a fixed form of process noise is not appropriate for this problem. Moreover, specification of  suitable noise parameters (i.e., $\sigma_{v}$ here) is often not obvious.

\begin{figure}[ptbh]
  \centering
 	\includegraphics[width=6in]{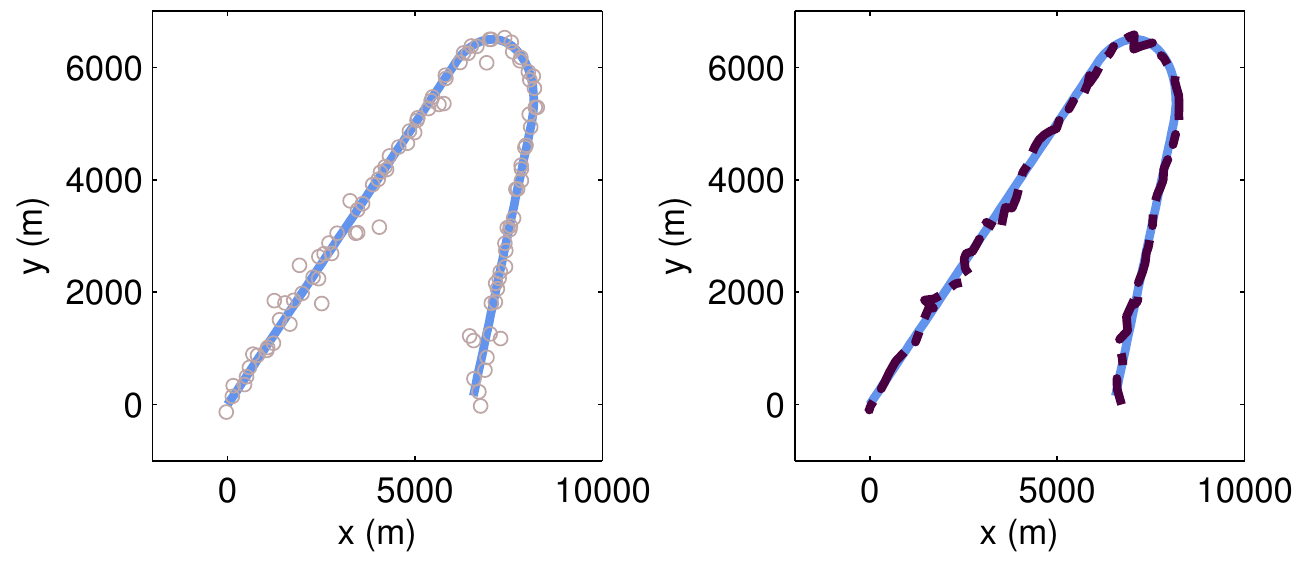}
  \caption{(left) true trajectory with measurements corrupted by outliers, (right) estimated trajectory by robust noise adaptive model}
  \label{fig:RBPFboth_target}
\end{figure}

Subsequently, we try our robust noise adaptive (RBPF) filter. Here the prior for the process noise is taken as a symmetric Laplace \big($p(v_k)\sim \mathcal{L}(0,10^6)$\big) and we use $50$ particles and $\rho = 0.05$. A typical estimated track is shown in Figure (\ref{fig:RBPFproc}). We also implement an IMM filter \cite{blomB:1988} with (Gaussian process noise) $\sigma_{v}=1$ and $\sigma_{v}=50$. We assume that the initial probability for the modes are equal and a mode transition probability matrix $\pi =\left[ 0.9 \ 0.1;\ 0.1 \ 0.9 \right]$. A typical estimate of the track is shown in Figure (\ref{fig:IMM}). We observed that both our RBPF and IMM are performing well for this problem. We also compare them over 20 Monte Carlo runs. The statistics for the maximum absolute errors (max abs err) and the average root mean square error (avg. RMSE) are shown in  Table \ref{tab:err_stat}.
\begin{table}[htbp]
	\centering
		\begin{tabular}{ l | c | c }\hline
		 & IMM & RBPF \\ \hline
		max abs error &  164.95& 191.87	\\ \hline
		avg. RMSE & 65.33 & 69.71 \\
		\hline
		\end{tabular}
	\caption{Error statistics over 20 Monte Carlo runs}
	\label{tab:err_stat}
\end{table}
From Table \ref{tab:err_stat}, we see that IMM is performing slightly better, although at the cost of requiring a properly specified mode transition matrix $\pi$. In practice, often $\pi$ is unknown and estimating $\pi$ online is an arduous task. This problem does not arise in our approach. Finally, there may be outliers in the measurement noise as well e.g., due to occasional sensor malfunctions. This problem cannot be solved trivially by tuning the parameters of the noises in a KF setup. It is hard for any filter to distinguish immediately whether any outlier is arising from process or measurement noise. To test our approach, we keep the same trajectory, but now the measurements are generated from a mixture distribution given by  $e_k\sim 0.8\mathcal{N}(0,diag[80^{2}, \ 80^{2}]) + 0.2\mathcal{N}(0,diag[300^{2}, \ 300^{2}])$. We implement our algorithm with priors for the process and the measurement noises to be Laplace and Students' \textit{t} noise respectively, given by $p(v_k)\sim \mathcal{L}(0,10^8)$ and $p(e_k)\sim \mathcal{T}(0,10^8;4)$. Again we use $50$ particles and $\rho = 0.05$. The filter is performing reasonably well in most cases. A typical estimate of the track is shown in  Figure (\ref{fig:RBPFboth_target}).     
%
	%
\section{Concluding remarks}\label{conclusion}
\noindent
This article considers the difficult online inference problems for linear dynamic systems under a very realistic set-up, where (a) the noises can have non-Gaussian densities (either appearing naturally or are modeled to accommodate outliers) and (b) the noise parameters can be unknown and time varying. The corresponding non-Gaussian density is characterized in terms of the skewness and/or the heavy tails, that is commonly observed in many practical applications of interest. We note that unlike the heavy tails, inference with the skewed noise has not attracted considerable research attentions. 

For the inference task, we envisage here a new Rao-Blackwellized particle filter by leveraging on a so called hierarchical Gaussian model on the noises. We showed that the proposed framework is not only robust to any noise outlier, but also adaptive  to potentially unknown and time varying noise parameters.  However, the framework requires a valid transition kernel for the intractable state, targeted by the particle filter. We  outlined how such kernels can be constructed, at least for certain classes of noises that cover many commonly occurring (non-Gaussian) noises,  where the well explored Students' \textit{t} is just a special case. The framework essentially runs a bank of (interacting) Kalman filters and so is relatively easy to understand and/or implement. We also explained how the 
problem can easily scale up with the dimensions using Rao-Blackwellization idea. The proposed algorithm here is very simple and flexible, although a bit computationally demanding. We subsequently carried out numerical experiments including a comparison to IMM. We observed empirically that  the framework is doing a very good job even with only $50$ particles. The associated tasks of (offline) state smoothing, model parameters learning and also extending the framework for space-time inference are left as future works.

%
	%
\section*{Acknowledgment}
\noindent The author would like to thank COOP-LOC, funded by SSF and
CADICS, funded by Swedish Research Council (VR) for the financial supports.
%
%


\bibliographystyle{IEEEtran}
\bibliography{IEEEabrv,bibtex_db}

\end{document}